\documentclass[fleqn,usenatbib]{mnras}

\usepackage{graphicx}	% Including figure files
\usepackage{amsmath}	% Advanced maths commands
\usepackage{amssymb}	% Extra maths symbols
\usepackage{multicol}        % Multi-column entries in tables
\usepackage{bm}		% Bold maths symbols, including upright Greek
\usepackage{pdflscape}	% Landscape pages
\usepackage{hyperref}
\usepackage{amsfonts}
\usepackage{txfonts}
\usepackage{ctable}
\usepackage{url, lineno}
\usepackage[normalem]{ulem}

\usepackage{color}
\usepackage{grffile}

\newcommand{\kms}{\,km\,s$^{-1}$} % kilometres per second
\newcommand\altaffilmark[1]{$^{#1}$}
\newcommand\altaffiltext[1]{$^{#1}$}
\newcommand{\driftvel}{{\bm w}}
\newcommand{\driftvelmag}{w}

\newcommand \beq        {\begin{equation}}
\newcommand \beqa	{\begin{eqnarray}}

\newcommand \eeq	{\end{equation}}
\newcommand \eeqa	{\end{eqnarray}}

\newcommand \HH	        {{\rm H}_2}

\newcommand{\dvel}{{\bm v}}
\newcommand{\gvel}{{\bm u}}

\newcommand{\drift}{{\bm w}}

\newcommand \pr {{Phys.~Rev.}}

\newcommand{\oldtext}[1]{}

\usepackage[T1]{fontenc}
\usepackage{ae,aecompl}

\title[Dust velocities]{Dust dynamics in \textsc{RAMSES} – II. Equilibrium drift velocity distributions of charged dust grains
 \vspace{-0.5cm}}

\author[Moseley \& Teyssier]{
\parbox[t]{\textwidth}{ 
	Eric R.~Moseley\altaffilmark{1}\thanks{E-mail: moseley@princeton.edu},
	R. Teyssier\altaffilmark{1}
} 
\vspace*{6pt} \\
\altaffiltext{1}{Department of Astrophysical Sciences, Princeton University, Princeton, NJ 08540, USA}\\
}

\date{}

\pubyear{2024}

\begin{document}
%\label{firstpage}
%\pagerange{\pageref{firstpage}--\pageref{lastpage}}
\maketitle
 % I accept that I do not choose the outcome, only the steps I take along the way. I take those steps with love.
% Abstract of the paper
\begin{abstract} We investigate the gas-grain relative drift velocity distributions of charged astrophysical dust grains in MHD turbulence. We do this using a range of MHD-PIC simulations spanning different plasma-$\beta$, sonic/Alfv{\'e}n Mach number, and with grains of varying size and charge-to-mass ratio. We find that the root-mean-square drift velocity is a strong function of the grain size, following a power law with a 1/2 slope. The r.m.s. value has only a very weak dependence on the charge-to-mass ratio. On the other hand, the shape of the distribution is a strong function of the grain charge-to-mass ratio, and in compressible turbulence, also the grain size. We then compare these results to simple analytic models based upon time-domain quasi-linear theory and solutions to the Fokker-Planck equation. These models explain qualitatively the r.m.s. drift velocity's lack of charge-to-mass ratio dependence, as well as why the shape of the distribution changes as the charge-to-mass ratio increases. Finally we scale our results to astrophysical conditions. As an example, at a length scale of one parsec in the cold neutral medium, 0.1 $\mu$m grains should be drifting at roughly 40\% of the turbulent velocity dispersion. These findings may serve as a basis for a model for grain velocities in the context of grain-grain collisions, non-thermal sputtering, and accretion of metals. These findings also have implications for the transport of grains through the galaxy, suggesting that grains may have non-negligible random motions at length-scales that many modern galaxy simulations approach. %These random motions may lead to both grain dynamical pressure as well as small scale diffusion, particularly along magnetic field lines.
\end{abstract}
\begin{keywords}
galaxies: ISM - ISM: kinematics and dynamics - ISM: dust, extinction - (magnetohydrodynamics) MHD - turbulence 
\end{keywords}
%\saythanks
\section{Introduction}
%\blindtext
Dust grains are important in determining the thermodynamics and chemistry of the interstellar medium (ISM). In addition to their absolute abundance, the grain size distribution is an important determinant of ISM chemistry and thermodynamics. Grain surfaces catalyze chemical reactions, such as the formation of $\HH$, an important ISM coolant \citep{Gould+Salpeter_1963}. Additionally, dust grains readily absorb radiation and re-emit that energy into the infrared, altering galaxy spectral energy distributions (SEDs) \citep{Desert+Boulanger+Puget_1990}. The chemical and thermodynamic properties of the ISM are thus affected not only by the absolute abundance of dust grains, but also by their size distribution. It is thus of interest how the size distribution evolves and how it comes about. There seems to be a relative deficit of small grains at redshift $z\gtrsim 6$ \citep{Aoyama+Hirashita_2020, Di_2021}. This could be because grains are produced initially at primarily super-micron sizes in the outflows of supernovae \citep{Todini+Ferrara_2001}. Presumably, these grains subsequently collide and produce a wide spectrum of smaller grains. However, the details of this process are poorly understood, especially when accounting for the fact that dust grains are charged. 

While the quantity directly relevant to grain-grain collisional processes is the grain-grain relative velocities, for grains of sufficiently different sizes, these velocities can be approximated by gas-grain relative velocities. This is because grain velocities become increasingly decorrelated as grain properties (charge-to-mass ratio, size) become increasingly disparate \citep{ormel2007closed, hirashita2012dust, hubbard2012turbulence}. Additionally, gas-grain relative velocities are important for understanding the non-thermal sputtering of grains, as well as grain accretion of metals.

Together with assumptions about the (in)dependence of gas-grain relative velocities and the underlying gas velocity, one may also use gas-grain relative velocities to understand the way that grains turbulently diffuse via small-scale random motions and resist compression via an effective dust ``pressure''. Determining a functional form for the dust velocity dispersion given turbulence and grain properties is equivalent to establishing a dust ``equation of state'' to close the moment hierarchy and thus establish dust equations of motion that can be effectively utilized across a wide range of scales, from the galactic down to the protoplanetary.

All of these things motivate a systematic survey of the gas-grain relative velocity distribution. Most of the work that has been done on the velocities of dust grains has been done in the context of protoplanetary discs (PPDs) and the formation of planetesimals. %\citep[e.g.][]{cuzzi2003blowing, pan2010relative, carballido2010relative}. %Summarize said work... 
This work has been conducted analytically \citep{volk1980collisions,cuzzi2003blowing, Yan+Lazarian+Draine_2004,ormel2007closed}, numerically \citep{dullemond2005dust,johansen2007rapid,hubbard2012turbulence}, and experimentally \citep{blum2008growth, guttler2010outcome}. These works have tended to characterize dust velocities with a single number (typically the r.m.s. velocity) rather than a distribution. Especially when considering dust-dust collisional processes, the distribution matters as much as the r.m.s., for with a distribution, there will generically be growth, bouncing, and shattering all occurring simultaneously. Depending upon the balance of these processes, the typical grain size will grow, shrink, or stagnate. In the process of growing from sub-micron to kilometer sizes, multiple processes impede growth, including the ``bouncing barrier'' \citep{zsom2010outcome}, ``fragmentation barrier'' \citep{blum2008growth}, ``drift barrier'' \citep{weidenschilling1977aerodynamics}, and ``charge barrier'' \citep{okuzumi2011charge}. Understanding these effects in detail requires an understanding both of typical grain velocities as well as grain velocity distributions.

Past work on this subject has focused on neutral grain species. This case is much simpler to deal with analytically. A common assumption about neutral species is that they assume velocities comparable to eddies with turnover times of order their stopping time \citep[e.g.][]{cuzzi2003blowing, hubbard2012turbulence}. However, this assumption breaks down with the introduction of charge, as more complex particle acceleration mechanisms are then permitted such as Fermi acceleration \citep{Fermi_1949}, gyroresonance \citep{Lazarian+Yan_2002}, shock acceleration \citep{guillet2007shocks}, time-transit damping \citep{Hoang+Lazarian+Schlickeiser_2012}, and others.

The work presented herein differs from and adds to previous work in several respects. As of this writing, it is the first and only study of its kind that systematically examines the way that the grain velocity distribution evolves using 3D MHD-PIC simulations as a function of turbulence and grain properties. Second of all, it is the only study that systematically examines the velocity \textit{distribution} of charged grains in MHD turbulence, although past work has examined the way that the r.m.s. gas-grain velocity is impacted by underlying turbulence \citep[e.g.][]{Yan+Lazarian+Draine_2004, Yan_2009, Hoang+Lazarian+Schlickeiser_2012}. Finally, the work we present here includes the supersonic and super-Alfv{\'e}nic regimes, another rarity in the literature. While work such as \citet{guillet2007shocks} examine the way that dust evolves in and impacts MHD shocks, the generality is limited by the one dimensional nature of the studies. This means that regions of interacting shocks and a full population of interstellar shocks are not included in these models.

In this paper, we investigate the way that the gas-grain relative velocity distribution changes with Mach number, plasma beta, grain charge-to-mass ratio, and grain size. In Section ~\ref{sec:methods}, we briefly recapitulate the methods outlined in \citet{moseley2023dust}, explain our conventions for dimensionless quantities, and explain the design of our simulations . In Section ~\ref{sec:analytic}, we explore a few novel analytic models for explaining the velocity distribution of grains. These include a Fokker-Planck model and a 1D time-domain quasi-linear model. Then, in Section ~\ref{sec:results}, we present the results of our numerical study. In Section ~\ref{sec:discussion}, we discuss the ways that our results may apply to physical ISM conditions. Finally, we summarize the paper and conclusions in Section~\ref{sec:summary}.

\section{Methods}\label{sec:methods}
Our numerical methods are described extensively in \citet{moseley2023dust}. We briefly recapitulate them here. 

We have built a magnetohydrodynamic-particle-in-cell (MHD-PIC) module on top of the astrophysical fluid code \textsc{RAMSES} \citep{teyssier2002cosmological, moseley2023dust}. This code treats dust as a collection of ``superparticles'', each of which represents an ensemble of grains with the same properties (size, charge, etc.). This method allows for full modeling of the underlying phase-space distribution. The dust particles modeled with this method are active, having mass, momentum, and energy that they may exchange with the surrounding gas; i.e. the particles have ``back-reaction''. 

All of the specific equations solved in this code are given in \citet{moseley2023dust}. The gas follows the standard equations of MHD combined with an isothermal equation of state and a dust back-reaction term. Reiterating the equation of motion for a single dust grain,
\begin{align}
    \frac{{\rm d}\dvel}{{\rm d}t} &= -\nu_{\rm s}(\dvel-\gvel) + \omega_{\rm L}(\dvel-\gvel)\times\hat{\bm b}, \label{eq:dustvel}
\end{align}
where $\frac{{\rm d}}{{\rm d}t}$ is the Lagrangian time derivative, $\dvel$ is the grain velocity, $\gvel$ is the fluid velocity, $\nu_{\rm s}$ is the stopping rate (the inverse of the stopping time $t_{\rm s}$, or roughly the time required for a grain to slow down by a factor of 2), $\omega_{\rm L}$ is the Larmor frequency, and $\hat{\bm b}$ is the unit vector along the direction of the magnetic field. 

In this study, we adopt the Epstein-Baines drag law, which is accurate in both sub-sonic and super-sonic particle velocity regimes \citep{epstein1924resistance,Baines+Williams+Asebiomo_1965, kwok1975radiation, Draine+Salpeter_1979a}. This drag law is given by,
\begin{align}
    \nu_{\rm s} &= \sqrt{\frac{8}{\pi\gamma}}\frac{\rho c_{\rm s}}{\rho_{\rm d}^{\rm i} a_{\rm gr}}\bigg(1+\frac{9\pi\gamma}{128}\frac{\drift^2}{c_{\rm s}^2}\bigg)^{1/2}.
    \label{eq:drag}
\end{align}
Where $\gamma$ is the gas adiabatic index, $\rho$ is the local gas density, $c_{\rm s}$ is the sound speed, $\rho$ is the gas density, ${\bm w}$ is the gas-grain relative velocity, $\rho_{\rm d}^{\rm i}$ is the grain internal mass density, and the grain radius is $a_{\rm gr}$. 

The method we use to advance dust velocities is operator split, fully implicit for the drag operator and semi-implicit for the Lorentz operator, allowing for no additional time-step constraints beyond those used in ordinary MHD, plus an additional constraint on how far a particle may move in a given time-step. This is true even in the limits where there is arbitrarily large concentration of solids $\mu \gg 1$ ($\mu$ is the dust-to-gas mass ratio) and arbitrarily small dust stopping and Larmor times. However, in practice, to maintain accuracy, we employ a time-step constraint so that there is always at least ten time-steps per the minimum particle Larmor period, ignoring effects from concentration of solids. 
% Define the parameters

\subsection{Units and dimensionless quantities}
In order to maintain as much generality as possible, we have opted to parameterize our survey by four dimensionless parameters, two of which are intrinsic grain properties and two of which are extrinsic gas properties. These are the grain size parameter $\varepsilon$, the charge-to-mass ratio parameter $\xi$, the sonic Mach number $\mathcal{M}$, and the Alfv{\'e}n Mach number $\mathcal{M}_{\rm A}$. We will sometimes refer to the plasma-$\beta$ rather than the Alfv{\'e}n Mach number, as the information contained in $\mathcal{M}$ and $\mathcal{M}_{\rm A}$ is equivalent to $\mathcal{M}$ and $\beta$ for an isothermal equation of state. Throughout this paper, all quantities are defined after the simulations have reached a steady state. 

These parameters are defined by,
\begin{align}
    \mathcal{M} &\equiv \sigma( \gvel)/c_{\rm s}, \\
    \mathcal{M}_{\rm A} &\equiv \sigma(\gvel)/v_{\rm A}  = \sigma(\gvel)\sqrt{4\pi\rho_0}/B_0,\\
    \mathcal{\beta} &\equiv \frac{8\pi P_0}{B_0^2} = \frac{8\pi \rho_0 c_{\rm s}^2}{B_0^2}.
\end{align}
$\sigma(\gvel)$ is the velocity dispersion of the gas, $c_{\rm s}$ is the isothermal sound speed, $B_0$ is the initial magnetic field strength, $\rho_0$ is the mean gas density, $v_{\rm A} = B_0/\sqrt{4\pi\rho_0}$ is the Alfv{\'e}n speed, and $P_0= \rho_0 c_{\rm s}^2$ is the mean pressure.

We will also at times refer to the transverse and parallel sonic Mach numbers, $\mathcal{M}_\bot$ and $\mathcal{M}_{||}$, as well as the transverse and parallel Alfv{\'e}n Mach numbers $\mathcal{M}_{{\rm A},\bot}$ and $\mathcal{M}_{{{\rm A},||}}$. These are computed from the r.m.s of those velocity components transverse to the mean magnetic field and the r.m.s. of the velocity component along the mean magnetic field, respectively. Both the Alfv{\'e}n and sonic Mach numbers are defined after simulations reach saturation, as simulations begin from an initially uniform, at rest medium.

The grain properties are defined by,
\begin{align}
    \varepsilon &\equiv \frac{\rho_{\rm d}^{\rm i} a_{\rm gr}}{\rho_{0}\ell_0},\\
    \xi &\equiv \omega_{\rm L}t_{\rm A} = \frac{Z e}{m c}\ell_0 \sqrt{4\pi \rho_0}.
\end{align}
%$\omega_{\rm L}$ is the Larmor frequency of a grain, $t_{\rm A}$ is the Alfv{\'e}n crossing time (time for an Alfv{\'e}n wave to cross the domain), $Z$ is the charge on a grain in units of the fundamental charge $e$, $m$ is the mass of a grain, and $c$ is the speed of light. 
$\varepsilon^{-1}$ and $\xi$ are analogous quantities, with $\varepsilon^{-1}$ quantifying the degree of coupling via drag and $\xi$ the degree of coupling via the Lorentz force. In the first line, $\varepsilon$ is roughly the grain stopping time (in the sub-sonic regime) divided by the sound-crossing time, and $\ell_0$ is a unit length scale (for us, always the linear extent of the simulation domain). In the second line, the charge-to-mass ratio parameter $\xi$ is given by the Larmor frequency $\omega_{\rm L}$ times the Alfv{\'e}n-crossing time $\ell_0/v_{\rm A}$, where $v_{\rm A}$ is the Alfv{\'e}n velocity. $Z$ is the grain charge in units of the fundamental charge $e$, $m$ is the grain mass, and $c$ is the speed of light.

\begin{figure*}
    \centering
    \includegraphics[width=0.8\textwidth]{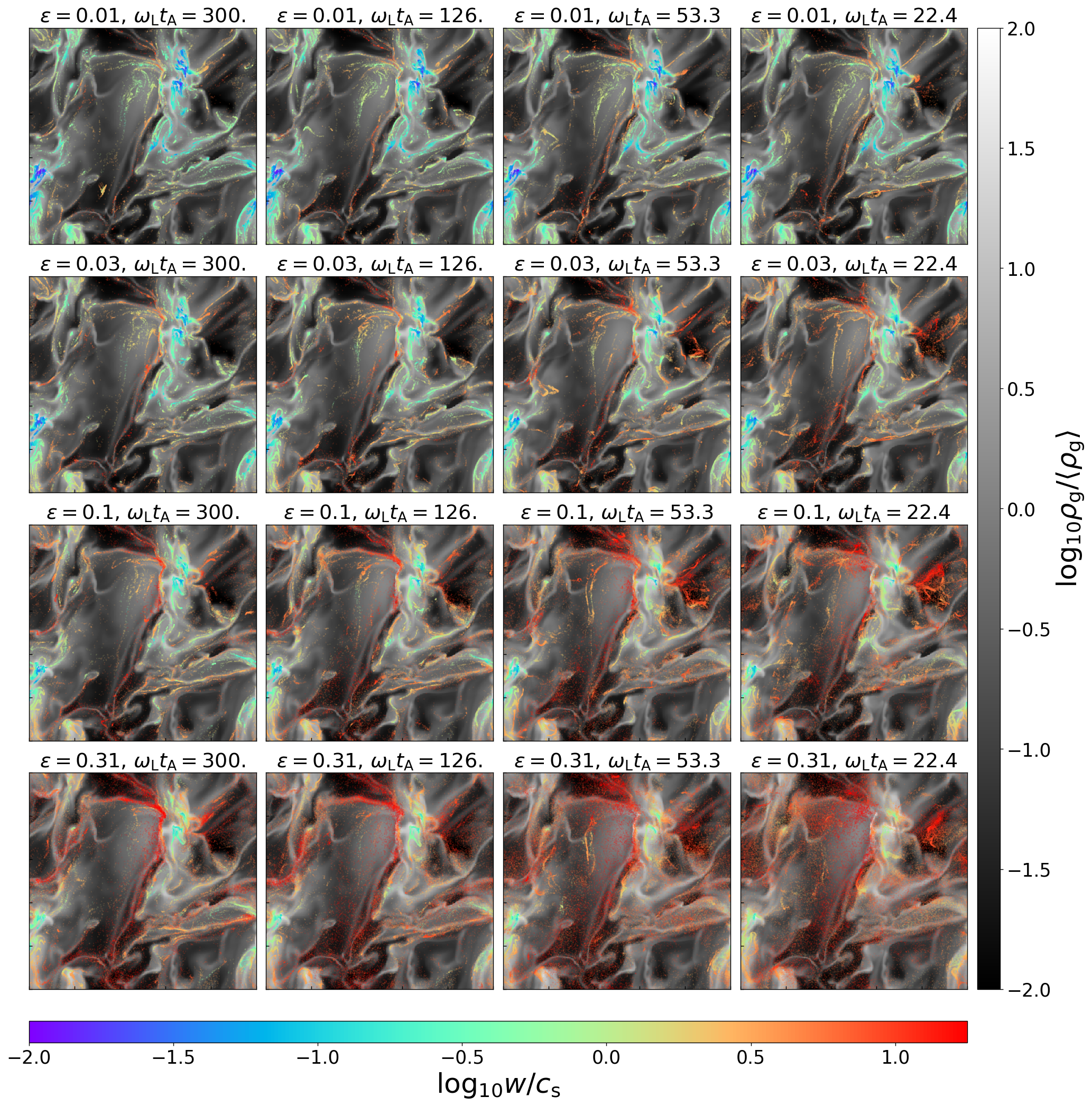}
    \caption{Above we have displayed a thin slice of the simulation b1m10 (cf. Tab.~\ref{tab:sims}), with plasma-$\beta$ equal to 1 and sonic Mach number approximately 9. The mean magnetic field points towards the top of the page. The background colors (going from black to white) represent the logarithm of gas density, $\rho_{\rm g}$, while the colored points represent the positions of dust superparticles. The dust superparticles are colored by the logarithm of the magnitude of the gas-grain drift velocity, $\driftvelmag$. Each plot is also labeled with the grain size parameter $\varepsilon$ as well as the charge parameter $\xi = \omega_{\rm L}t_{\rm A}$, which is equivalent to the Larmor frequency times the Alfv{\'e}n crossing time. It is evident here that for more coupled grains (lower $\varepsilon$, higher $\xi$), their typical drift velocities are lower. It is also clear that the denser the gas, the more strongly coupled the grains are. In these panels is also a clear instance of shock acceleration, near the center of the plots; a complex shock front is rimmed with an overdensity of high velocity grains, appearing as a red region tracing the shock front. The most dramatic case is the bottom left panel, with the weakest drag and the highest charge-to-mass ratio. One may also notice how more weakly coupled grains are more dispersed throughout the medium, while more strongly coupled grains cluster more strongly into dense regions.}
    \label{fig:grid}
\end{figure*}

\subsection{Simulation design}
We have run a range simulations at a resolution of $256^3$ gas cells and $16 \times 256^3$ dust particles at a dust-to-gas mass ratio of $\mu = 0.01$ and an isothermal equation of state. Within each simulation, we have 16 different dust species, each with $256^3$ particles. These simulations are similar to the decaying turbulence simulations we presented in \citet{moseley2023dust}, but with multiple dust species rather than just one, and driven continuously (to saturation) rather than decaying. The grain species span a range of charge-to-mass ratio and grain size, while the simulations themselves span a range of plasma beta and turbulent sonic Mach number (detailed in Tab.~\ref{tab:sims}). In particular, within each simulation, there are four possible values for the charge-to-mass ratio parameter $\xi_i$ and 4 possible values for the grain size $\varepsilon_j$. We then take all different combinations of these, to get 16 different combinations of $(\xi_i,\varepsilon_j)$. This allows us to consider the impact of varying one of these two parameters while the other is held fixed within each simulation. We design this study around these parameters because these correspond directly to quantities upon which there are resolution requirements, the Larmor radius $r_{\rm L}$ and the grain stopping length $\ell_{\rm s}$ (the distance for a grain to slow down by a factor of $~2$). Additionally, these are coefficients that quantify directly the strength of the terms in Eq.~\ref{eq:dustvel}. Fig.~\ref{fig:grid} shows a thin slice of a simulation (b1m10, Tab.~\ref{tab:sims}) that highlights the design of this study, showing dust superposed on top of gas. The dust is colored by the drift velocity, while the gas is colored by density. 

Again, grains in these simulations are not passive, but fully active particles. For a dust-to-gas mass ratio of 0.01, dust-dust collective effects (where each species may impact the other) are of order $\mu^2 = 10^{-4}$, and so are ignorable. However, for the highest velocity grains in the simulation, dust-gas momentum feedback may not necessarily be negligible, so we include this effect for completeness at the expense of only a modestly higher computational cost.

We mostly restrict ourselves to the range where $r_{\rm L} \equiv w_\bot/\omega_{\rm L}$ and $\ell_{\rm s} \equiv w t_{\rm s}$ are in the approximate range $(0.01\ell_0,1\ell_0)$, with a few grain populations falling outside this range. This ensures that for any effects where grain back-reaction is important, the dynamics are appropriately resolved. If grains have an $r_{\rm L}$ or $\ell_{\rm s}$ which falls below the simulation's numerical viscous scale, we expect that grain velocities will be suppressed by an additional amount \citep{Yan+Lazarian+Draine_2004}. The other limit, where $r_{\rm L}$ or $\ell_{\rm s}$ are greater than $\ell_0$ is more subtle. While not necessarily unphysical, in this limit, grains are effectively decoupled from the gas, with grains only weakly perturbed over the course of a turbulent turnover time. If we imagine we are simulating individual ISM clouds, then grains would entirely leave the cloud during this time. We may instead imagine instead that this case represents simply very small scales within larger clouds. We also must run the simulation for at least a few multiples of ${\rm max}(t_{\rm s},2\pi/\omega_{\rm L},t_{\rm dyn})$ ($t_{\rm dyn}$ is the turbulent turnover time) in order to reach statistical equilibrium.

Our simulations span (roughly) three different regimes for each of Mach number and plasma-$\beta$, with each roughly taking on values of 0.1, 1, and 10. In total, we have run 10 simulations. We run our simulations to saturation and beyond, each for roughly 8 turnover times. Thus, we span four independent parameters, plasma-$\beta$, sonic Mach number $\mathcal{M}$, grain charge-to-mass ratio parameter $\xi$, and grain size $\varepsilon$ with 160 different dust species across 10 different simulations. Simulations are driven with a compressive fraction $\chi = 0.9$ using an Ornstein-Uhlenbeck process with a correlation time set to be approximately the expected turbulent turnover time. $\chi$ is defined using the standard Helmholtz decomposition \citep[e.g.][]{federrath2008density}, with $\chi=0$ being a vector field that is divergence-free and $\chi=1$ being a vector field that is curl-free. We choose $\chi=0.9$ because sources of turbulence in the real ISM are expected to be more compressive than solenoidal \citep[e.g. supernovae, gravity;][]{mac2004control, elmegreen2004interstellar}.

We also employ an additional time-step constraint beyond the standard Courant condition so that $\omega_{\rm L}\Delta t/(2 \pi) < 0.05$ is true for all grains in the simulation. While not necessary for stability, this ensures that all grain gyro-orbits are well-resolved in time. The only exception to this rule is in the case of strong concentration of solids, especially of a single grain species. In this case, it is possible that collectively grains oscillate at a higher frequency, but due to the implicit algorithm we employ, this (rare) event should not yield unphysical results. For a single grain species, the natural oscillation frequency of the system is $(1+\mu)\omega_{\rm L}$ rather than $\omega_{\rm L}$. This is explained in Sec. 3.1 of \citet{moseley2023dust}. We employ no additional constraint for the drag, as it is almost always the case that $\omega_{\rm L}t_{\rm s} > 1$ for the grains we have modeled. 

We also use the HLLD Riemann solver together with one of two slope limiting procedures: subsonic simulations use the MonCen slope limiter, while supersonic and transsonic simulations use the MinMod slope limiter. 
\begin{table*}
\centering
\begin{tabular}{||c | c c | c c c | c c | c c||} 
 \hline
name & $\mathcal{M}_\bot$ & $\mathcal{M}_{||}$ & $\beta$ & $\mathcal{M}_{{\rm A},\bot}$ & $\mathcal{M}_{{\rm A},||}$ & $\varepsilon_{\rm min}$ & $\varepsilon_{\rm max} $& $\xi_{\rm min}$ & $\xi_{\rm max}$ \\ [0.5ex]
\hline
b10m0.1 & 0.10 & 0.023 & 10 & 0.22 & 0.052 & 1.00 & 13.33 & 7.50  & 100\\
b10m1 & 0.64 & 0.50 & 10 & 1.43 & 1.12 & 0.06 & 1.90 & 7.50 & 100 \\
b10m10 & 7.65 & 5.81 & 10 &  17.11 & 13.00 & 0.01 & 0.32 & 37.49 & 500 \\
\hline 
b1m0.1 & 0.10 & 0.016 & 1 &  0.073 & 0.011 & 1.00 & 13.33 & 2.37  & 31.62 \\
b1m1 & 0.88 & 0.41 & 1 & 0.63 & 0.29 & 0.06 & 1.90 & 2.37 & 31.62 \\
b1m1L & 0.93 & 0.40 & 1 & 0.66 & 0.29 & 0.6 & 19.0 & 2.37 & 31.62 \\
b1m10 & 7.11& 5.81 & 1 & 5.03 & 4.11 & 0.01 & 0.32 & 22.50 & 300 \\
\hline 
b0.1m0.1 & 0.10 & 0.015 & 0.1 &  0.023 & 0.0034 & 1.0 & 13.33 & 0.75  & 10 \\
b0.1m1 & 1.37 & 0.43 & 0.1 & 0.31 & 0.096 & 0.06 & 1.90 & 0.75 & 10 \\
b0.1m10 & 6.27 & 5.71 & 0.1 & 1.40 & 1.28 & 0.01 &  0.32 & 7.50 & 100 \\ [1ex] 
 \hline
\end{tabular}
\caption{List of parameters and statistics from simulations presented in this study. Each row is a single simulation. Each column is for a different parameter, namely the perpendicular and parallel Mach numbers $\mathcal{M}_\bot$, $\mathcal{M}_{||}$, the plasma-$\beta$, the perpendicular and parallel Alfv{\'e}n Mach numbers $\mathcal{M}_{{\rm A},\bot}$, $\mathcal{M}_{{\rm A},||}$, the minimum and maximum grain size parameters $\varepsilon_{\rm min} $, $\varepsilon_{\rm max} $, and the minimum and maximum charge parameters $\xi_{\rm min}$, $\xi_{\rm max}$. Directions parallel versus perpendicular refer to orientation of the velocity relative to the mean magnetic field.}
\label{tab:sims}
\end{table*}
\begin{figure*}
    \centering
    \includegraphics[width=\textwidth]{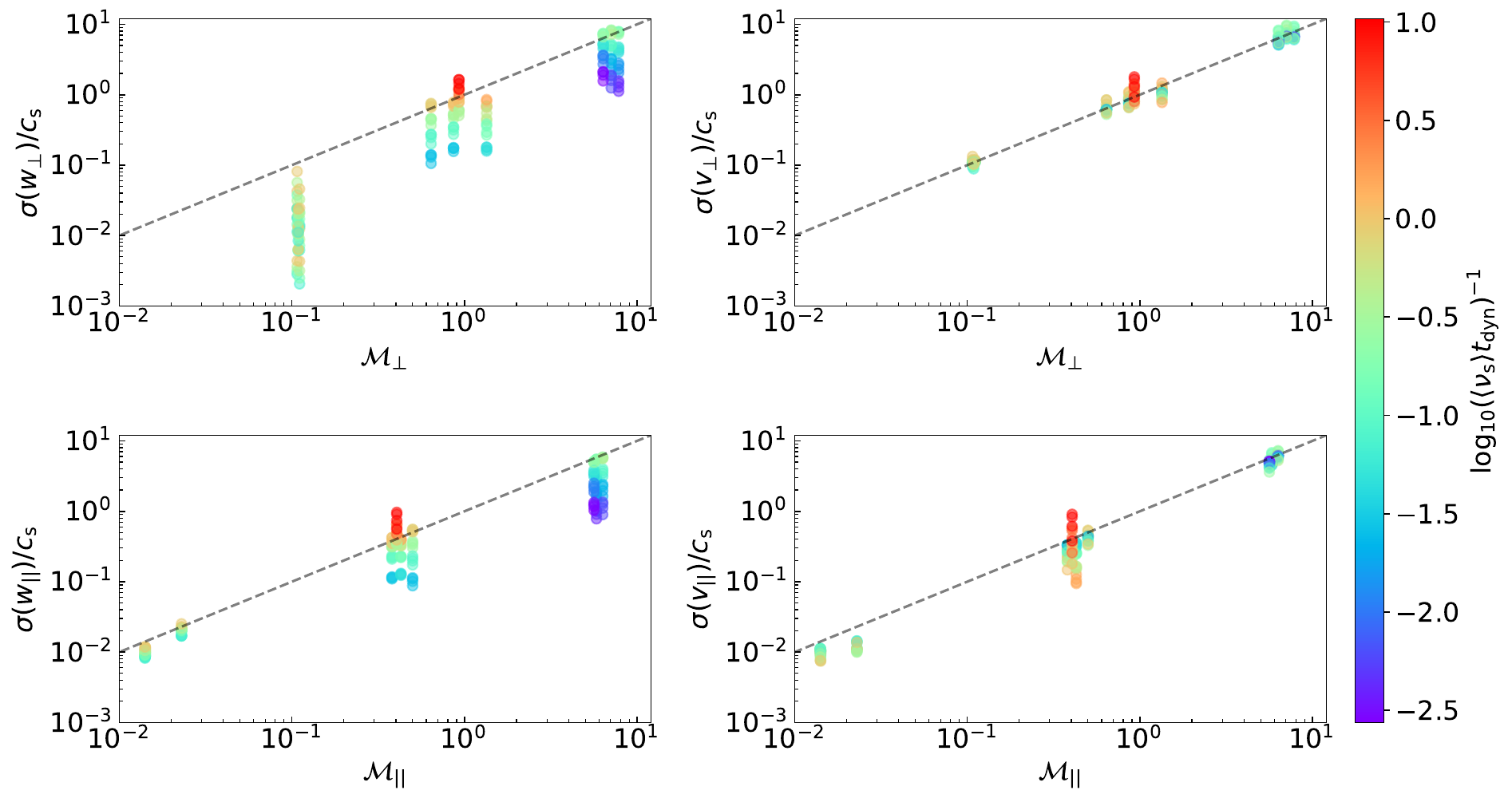}
    \caption{\textbf{Left:} Root-mean-square parallel and perpendicular dust drift velocities versus parallel and perpendicular sonic Mach number, respectively. Colors indicate one over the ensemble-averaged grain stopping rate for each species $\langle \nu_{\rm s} \rangle$ ($t_{\rm s}^{-1}$) times the dynamical time. This quantity is also known as the Stokes number. It roughly quantifies the strength of the grain drag force. Dashed lines indicate $\sigma(\drift) = \sigma(\gvel)$. \textbf{Right:} Root-mean-square parallel and perpendicular dust velocities vs parallel and perpendicular sonic Mach number, respectively. The colors are the same as the two panels on the left. Dashed lines indicate $\sigma(\dvel)= \sigma(\gvel)$. While both the velocity and the drift velocity dispersions of grains scale roughly with the gas velocity dispersion, the deviations from this behavior can be significant depending on the grain properties. As well, as MHD turbulence can be highly anisotropic, so too can the grain velocity and drift velocity distributions.}
    \label{fig:all_vels_vs_mach}
\end{figure*}
%we discuss observational implications as well as applications to theories of galactic dust evolution.
% \begin{figure*}
%     \centering
%     \includegraphics[width=\textwidth]{figures/deviation.pdf}
%     \caption{\textbf{Left:} Root-mean-square parallel and perpendicular dust drift velocities normalized by parallel and perpendicular gas velocity dispersions, respectively verses the ensemble-averaged grain Larmor frequency times the dynamical time. Stars show supersonic simulations, triangles show transsonic simulations, and squares show subsonic simulations. Colors indicate the ensemble-averaged drag coefficient times the dynamical time. \textbf{Right:} The same as the left, but for grain velocities rather than drift velocities. }
%     \label{fig:deviation}
% \end{figure*}

\begin{figure*}
    \centering
    \includegraphics[width=\textwidth]{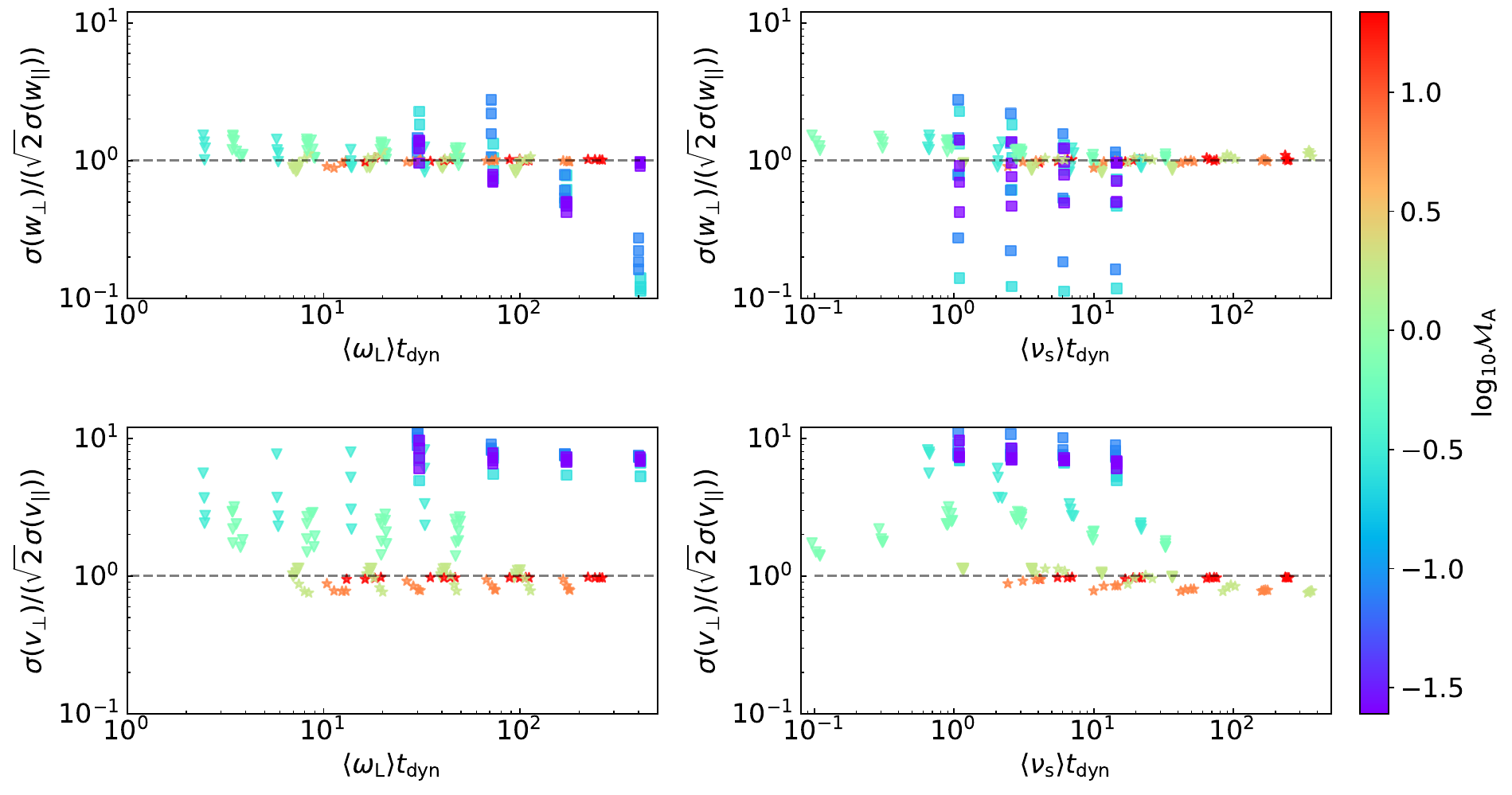}
    \caption{\textbf{Top Left:} Root-mean-square perpendicular dust drift velocities divided by the parallel root-mean-square drift velocities times the square root of two (to account for degrees of freedom) versus the ensemble-averaged Larmor frequency times the dynamical time. Colors indicate the Alfv{\'e}n Mach number. Stars show supersonic simulations, triangles show transsonic simulations, and squares show subsonic simulations. \textbf{Top Right:} The same, but instead as a function of the ensemble-averaged drag coefficient times the dynamical time. \textbf{Bottom Left:} The same as the top left, but with dust velocities rather than drift velocities. \textbf{Bottom Right:} The same as the top right, but with dust velocities rather than drift velocities.}
    \label{fig:asymmetry}
\end{figure*}

\begin{figure*}
    \centering
    \includegraphics[width=\textwidth]{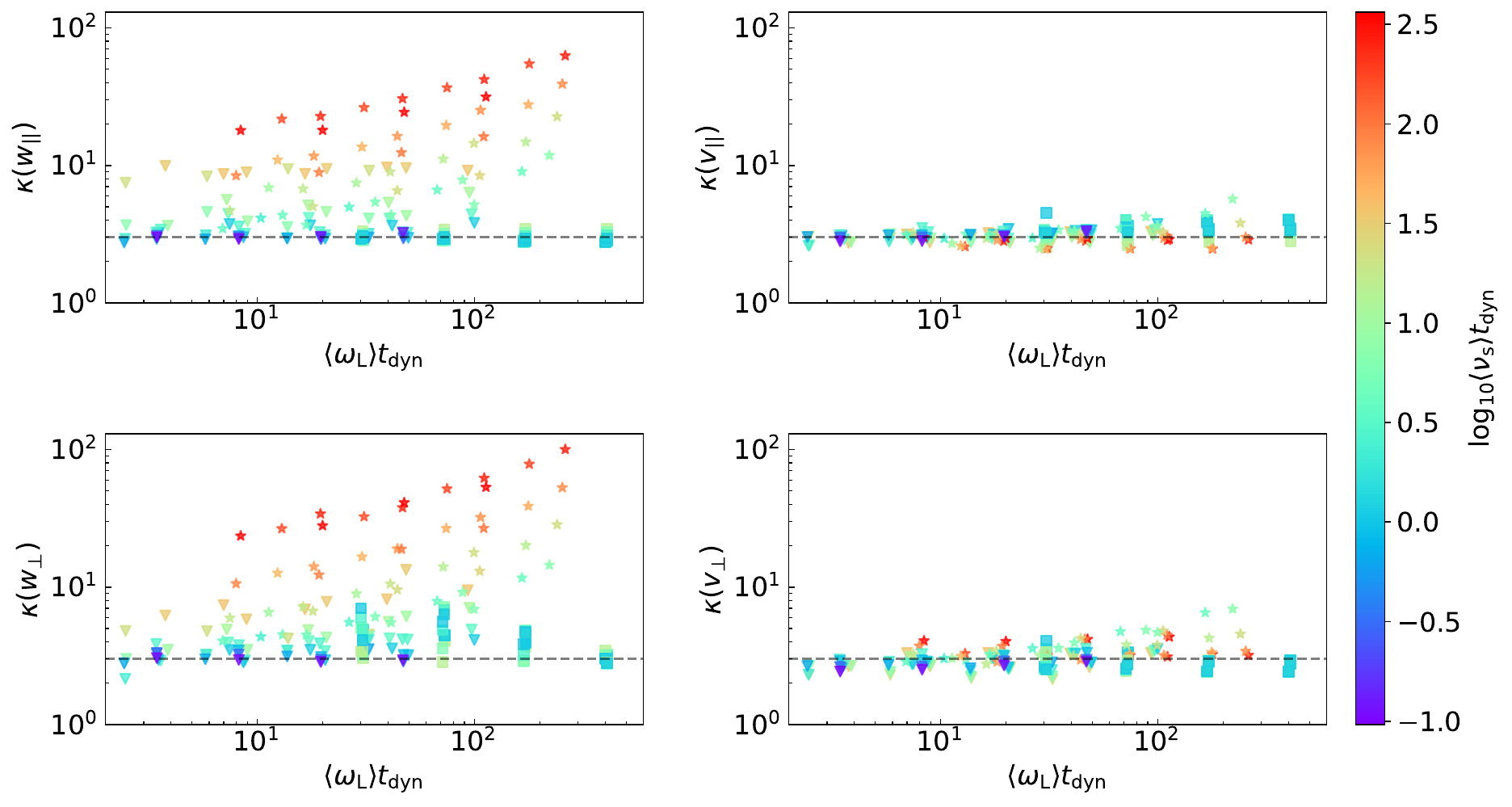}
    \caption{\textbf{Left:} The parallel (top) and perpendicular (bottom) kurtosis of the dust drift velocity distributions as a function of ensemble-averaged Larmor frequency times the dynamical time. Different point styles are the same as described in Fig.~\ref{fig:asymmetry}. Colors indicate the ensemble-averaged drag coefficient times the dynamical time. The horizontal dashed line is a kurtosis of three, which is the kurtosis of the normal distribution. \textbf{Right:} The same, but for the dust absolute velocities rather than drift velocities.}
    \label{fig:kurtosis}
\end{figure*}

\begin{figure*}
    \centering
    \includegraphics[width=\textwidth]{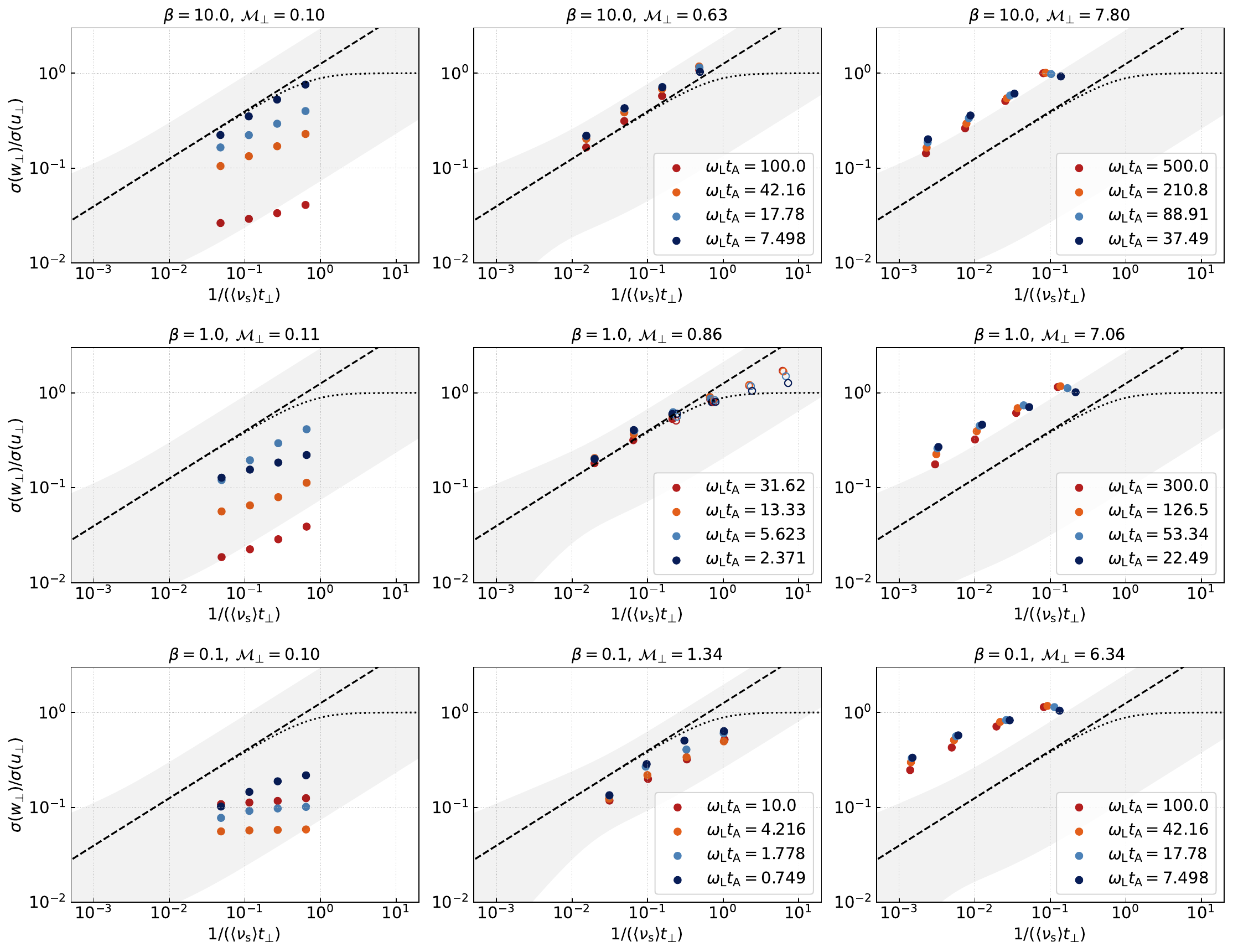}
    \caption{The grain drift velocity dispersion across the mean magnetic field direction normalized by the gas velocity dispersion in the same direction vs the mean stopping time of each population of grains in units of the perpendicular dynamical time $t_\bot = \ell_0\sqrt{2}/\sigma(\gvel_\bot)$. The leftmost column represents subsonic simulations, the center column transsonic simulations, and the rightmost column supersonic simulations. The top row shows $\beta=10$, the middle $\beta = 1$, and the bottom $\beta = 0.1$. On each plot, we show the analytic models discussed in Sec.~\ref{sec:qlt} for a $-5/3$ spectral index and charged grains (dashed line) as well as neutral grains (dotted line). The upper limit of the grey region is the same model but for a $-3/2$ spectral index and assuming that the charge is equal to that of the most strongly charged grains in the simulation, while the lower limit of the grey region is the same, but for a $-2$ spectral index. Each panel features 16 (or 32) different grain populations with different charge-to-mass ratio parameters $\omega_{\rm L}t_{\rm A}$ as indicated by the legend. In the case of the leftmost plots, their $\omega_{\rm L}t_{\rm A}$ are the same as the middle plots. In the center panel, we have plotted an additional simulation (indicated by open circles) that was run identically, but with grains effectively ten times larger (ten times weaker drag). This allows us to see the trend for grains where $t_{\rm s} > t_{\rm dyn}$. }
    \label{fig:wperp_vs_ts}
\end{figure*}

\begin{figure*}
    \centering
    \includegraphics[width=\textwidth]{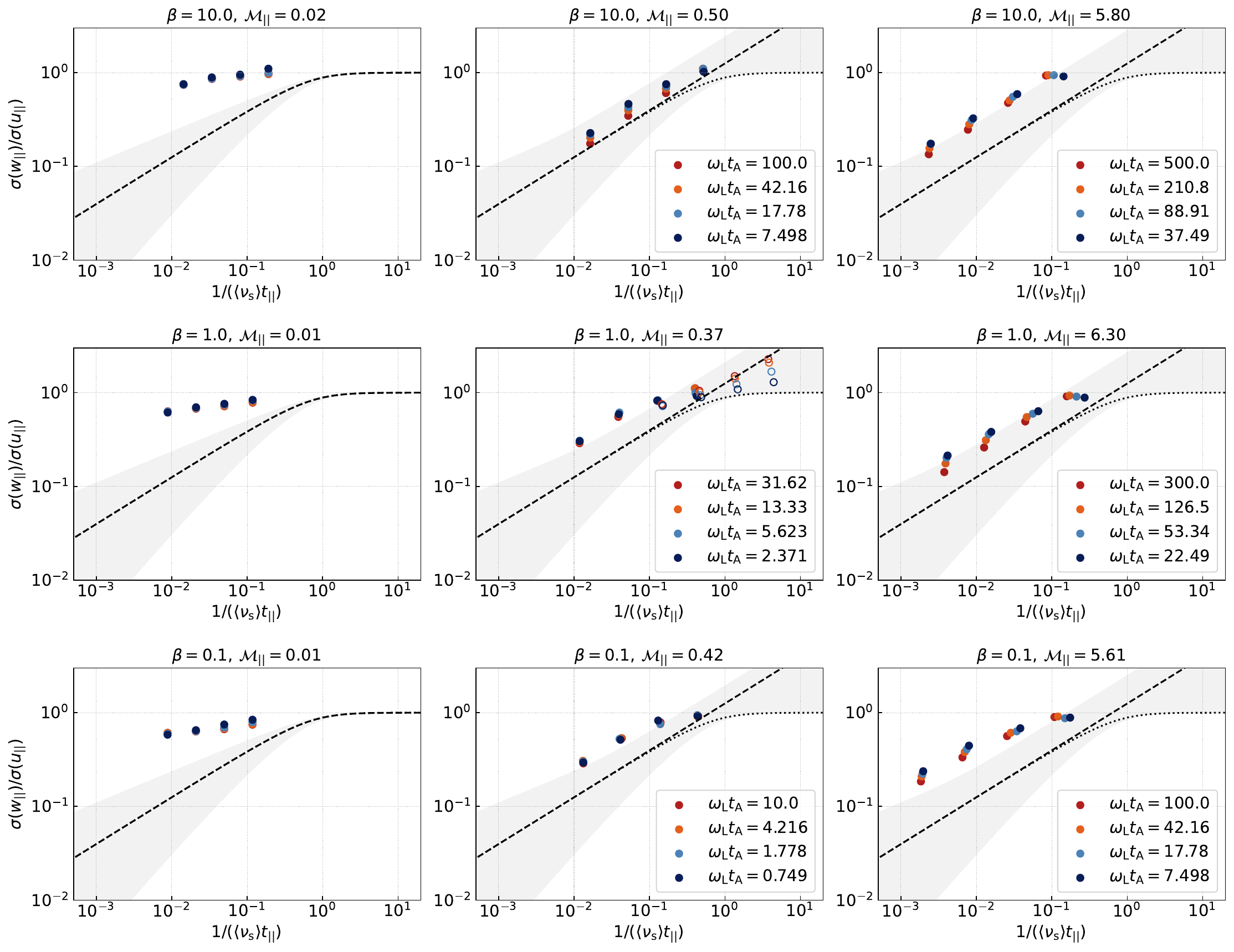}
    \caption{The same as Fig.~\ref{fig:wperp_vs_ts}, but for the direction along the mean field. The models shown by the dotted and grey boundaries in the leftmost panel assume neutral grains because the field is so relatively strong in that case that we expect charge to have little effect on the velocities; that is, acceleration processes become inefficient in the limit of a perfectly rigid field. In the other panels, the upper model assumes the largest charge-to-mass ratio parameter $\omega_{\rm L}t_{\rm A}$ present in the simulation, the dashed model is charge independent so long as $\omega_{\rm L}t_{\rm dyn} > 1$, and the lower model assumes neutral grains and a $-2$ spectral index. As with Fig.~\ref{fig:wperp_vs_ts}, the upper extent of the boundary is for a $-3/2$ slope, although for this direction we have also included a $-2$ model, indicated by the lower grey boundary. This model (applicable only for neutral grains, theoretically) is also shown  }
    \label{fig:wpara_vs_ts}
\end{figure*}

\begin{figure*}
    \centering
    \includegraphics[width=\textwidth]{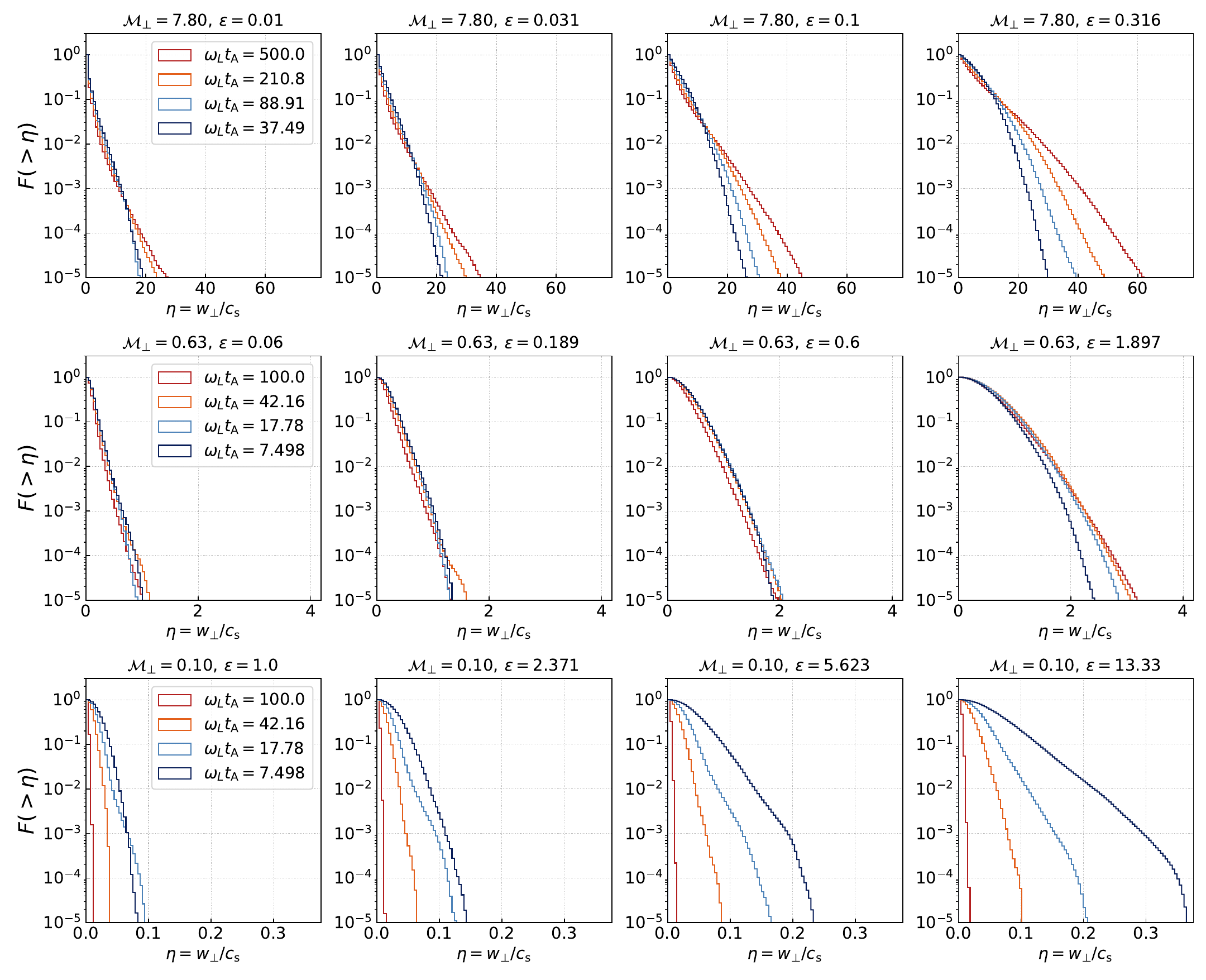}
    \caption{Drift velocity survival functions (SFs) perpendicular to the mean magnetic field for ${\beta_0=10}$. Each row shows the same simulation, while each column shows different grain-size parameters $\varepsilon$. Different colors indicate different charge-to-mass ratio parameters, here quantified in a dimensionless way as $\omega_{\rm L} t_{\rm A}$, the Larmor frequency times the Alfv{\'e}n crossing time. Each plot is also labeled by its respective sonic Mach number and grain size parameter (proportional to the grain size).}
    \label{fig:b10_perp}
\end{figure*}

\begin{figure*}
    \centering
    \includegraphics[width=\textwidth]{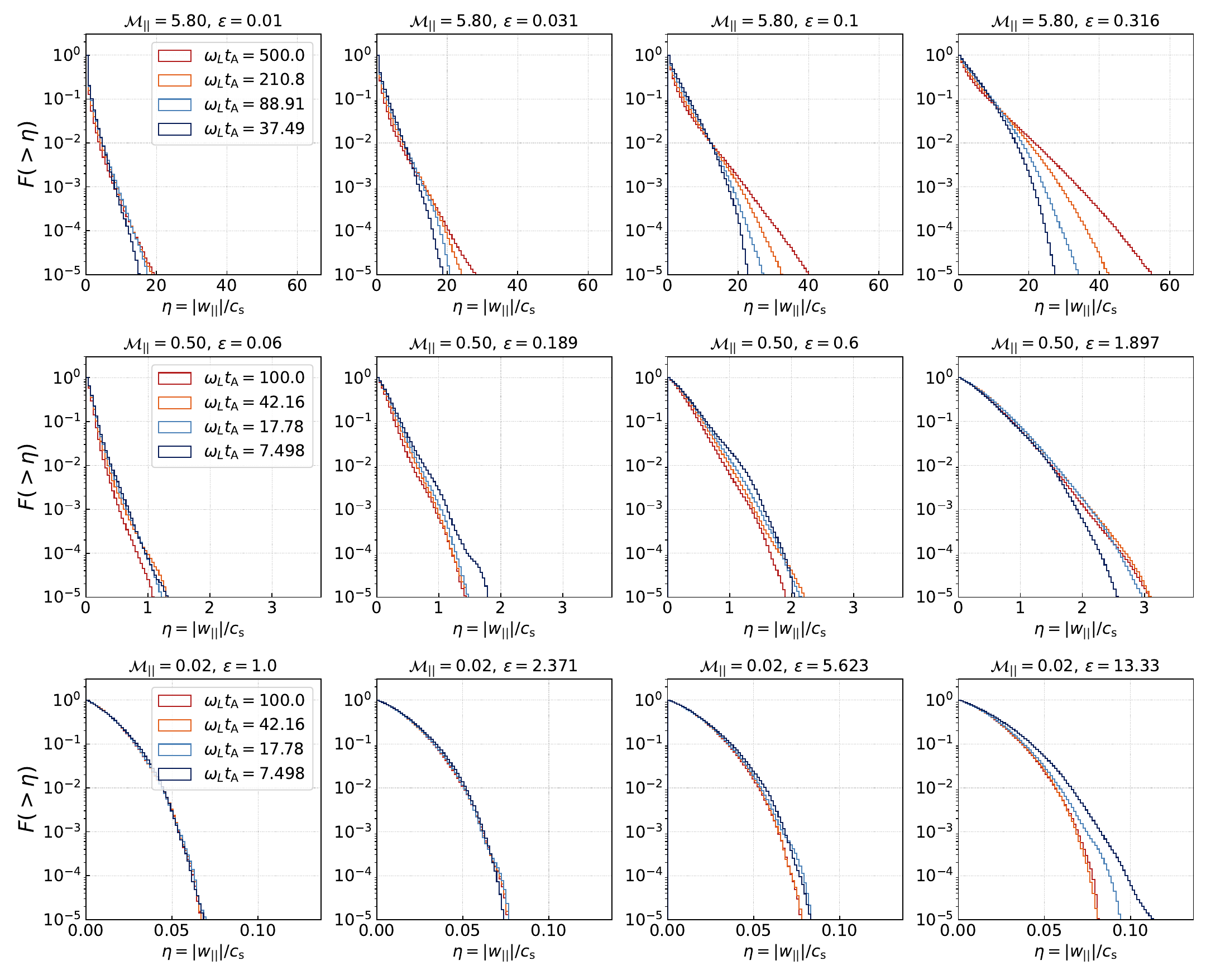}
    \caption{The same as Fig.~\ref{fig:b10_perp}, (so ${\beta_0=10}$ here as well) but for the direction parallel to the mean field. }
    \label{fig:b10_para}
\end{figure*}

\begin{figure*}
    \centering
    \includegraphics[width=\textwidth]{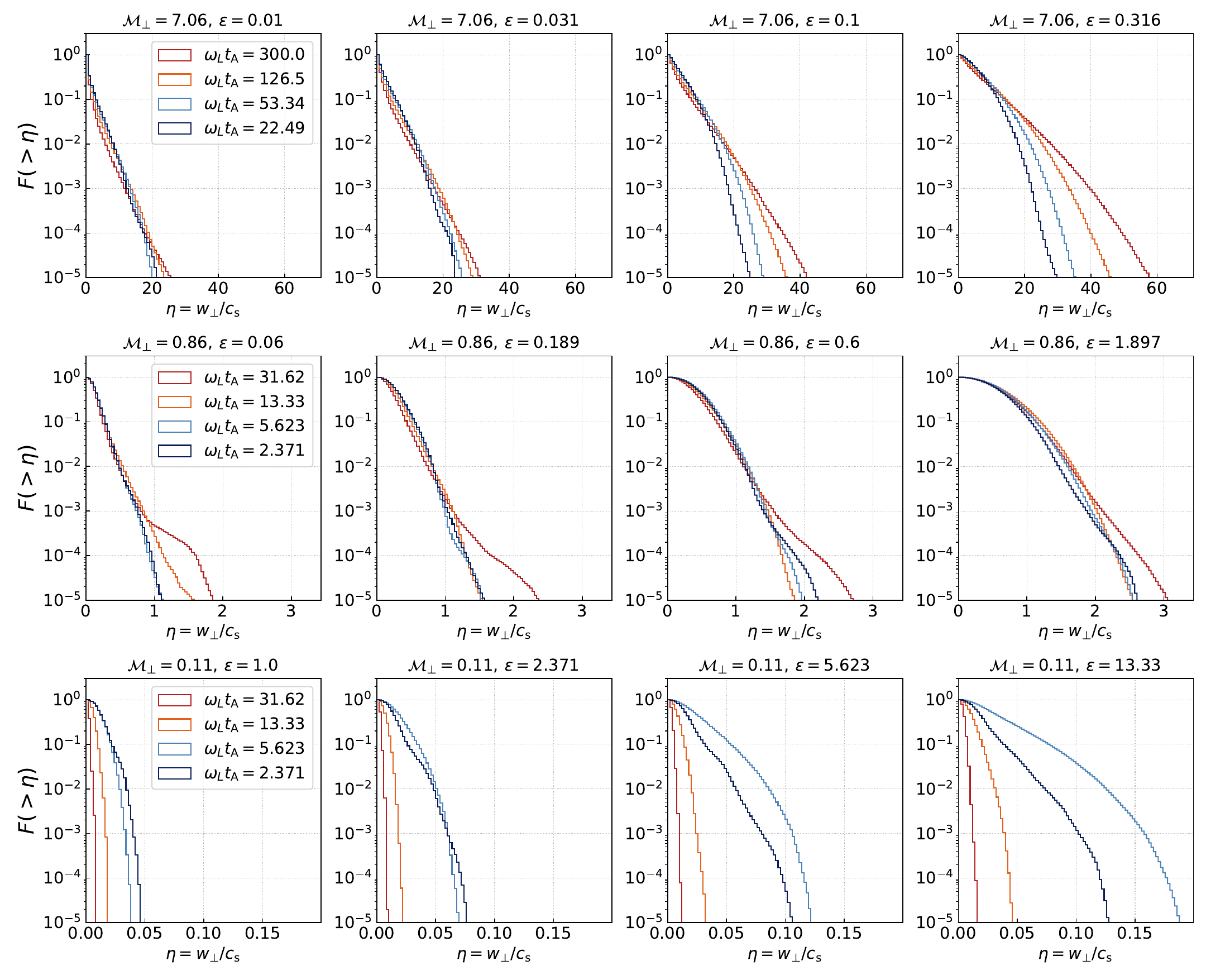}
    \caption{The same as Fig.~\ref{fig:b10_perp}, but for ${ \beta_0=1}$.}
    \label{fig:b1_perp}
\end{figure*}

\begin{figure*}
    \centering
    \includegraphics[width=\textwidth]{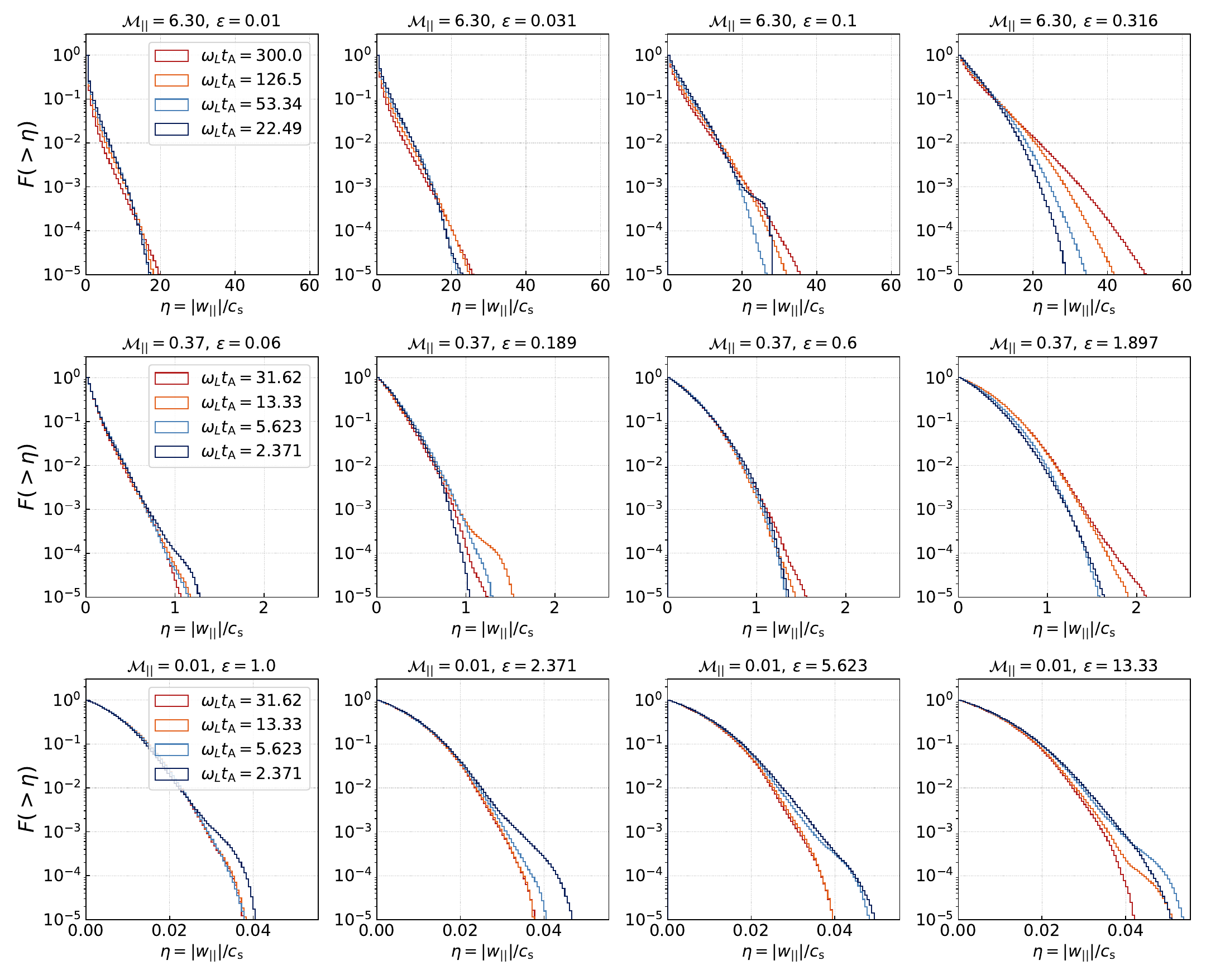}
    \caption{The same as Fig.~\ref{fig:b10_para}, but for ${ \beta_0=1}$. }
    \label{fig:b1_para}
\end{figure*}

\begin{figure*}
    \centering
    \includegraphics[width=\textwidth]{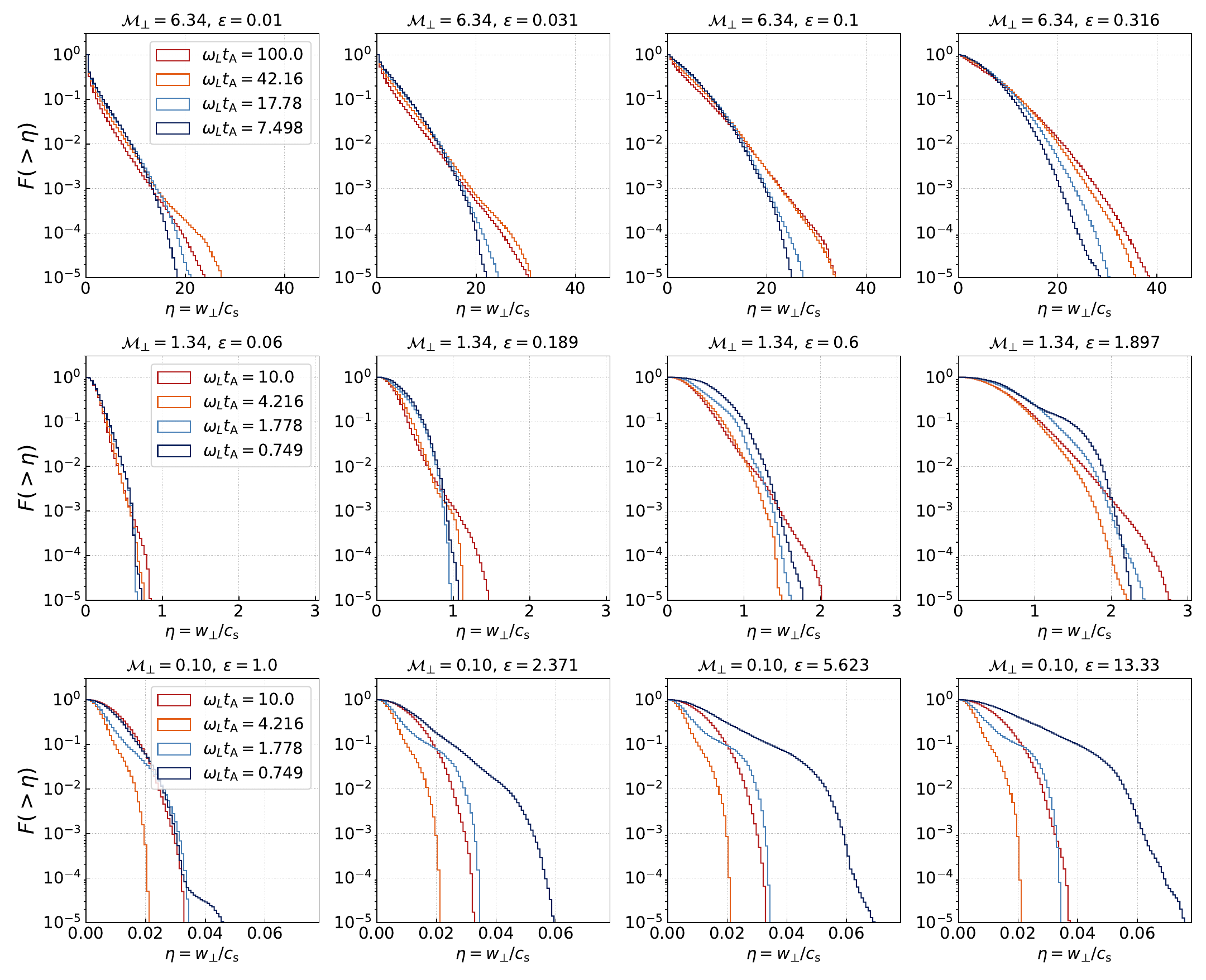}
    \caption{The same as Fig.~\ref{fig:b10_perp}, but for ${ \beta_0=0.1}$.}
    \label{fig:b1d-1_perp}
\end{figure*}

\begin{figure*}
    \centering
    \includegraphics[width=\textwidth]{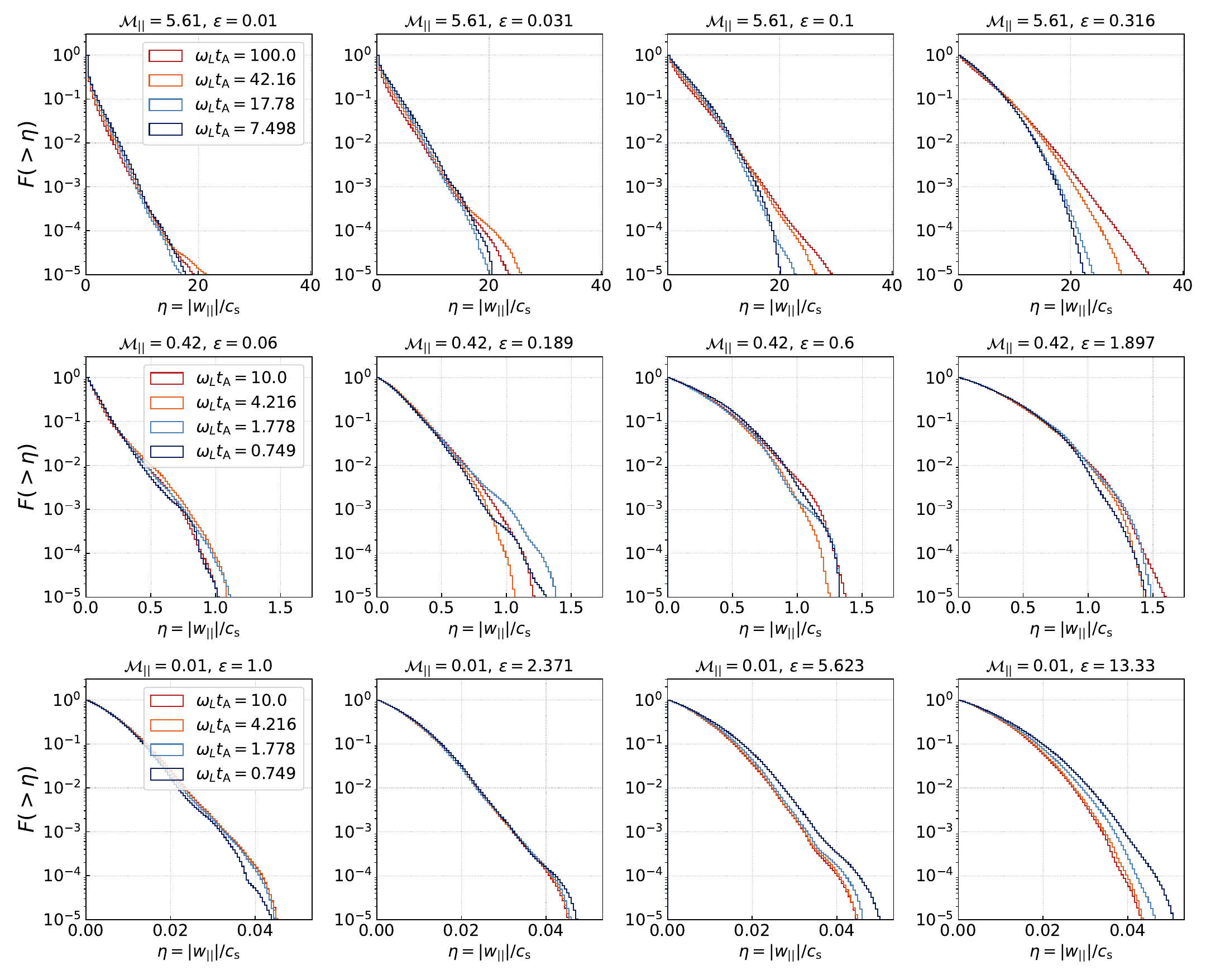}
    \caption{The same as Fig.~\ref{fig:b10_para}, but for ${ \beta_0=0.1}$.}
    \label{fig:b1d-1_para}
\end{figure*}

\section{Analytic models} \label{sec:analytic}
Below, we present a few simple analytic models to aid in our interpretation of the simulation data. While none of these models appears to be quantitatively correct, they do seem to capture qualitative features and trends in the data. 

\subsection{Grain motions as a time-domain-only process}\label{sec:qlt}
Eq.~\ref{eq:dustvel} gives the equation of motion for a dust grain, but to understand the evolution of the gas-grain drift velocity directly, we must instead work with the following equation:
\begin{align}
    \frac{{\rm d}}{{\rm d}t}\bigg|_\dvel\drift &= -(\nu_{\rm s}+\omega_{\rm L}\hat{\bm b}\,\times\,) \, \drift -\frac{{\rm d}}{{\rm d}t}\bigg|_\dvel \gvel. \label{eq:langevin}
\end{align}
${\rm d}/{\rm d}t|_\dvel$ is the convective derivative with the dust velocity $\dvel$. 

\subsubsection{Neutral grains}

For the purposes of comparison to previous work as well as to aid in understanding the impact of charge, we first present a model for neutral grains.

%We may now make the substitution $\hat{\bm b}\,\times \rightarrow i $ as in \citet{moseley2023dust}. This substitution combines the two components perpendicular to the mean field into a single complex number, e.g. $\dvel_\bot = v_x + i v_y$. This works so long as the field is perfectly rigid so that we may treat the perpendicular and parallel directions as independent from one another. 
%We now perform our analysis in the frame of the particle in the absence of the Lorentz force.  

Taking the Fourier transform of Eq.~\ref{eq:langevin} and allowing $\omega_{\rm L}=0$ gives,
\begin{align}
    -i \Omega \tilde{\bm w} &= -\nu_{\rm s} \tilde{\bm w} + i \Omega \tilde{\bm u},\\
    \implies |\tilde{{\bm w}}|^2 &= \frac{\Omega^2 t_{\rm s}^2}{1+\Omega^2t_{\rm s}^2}|\tilde{\bm u}|^2,
\end{align}
where $\Omega$ is the frequency of gas velocity fluctuations in the frame of a dust grain. In this case, theories on MHD turbulence such as those of \citet{goldreich1995toward} and \citet{boldyrev2005spectrum} tell us about $|\tilde{\gvel}|^2$.\footnote{This is only approximately true, due to the fact that the dust grain is never perfectly coupled or perfectly decoupled from the fluid. This means that the grain experiences a Doppler shift in the frequencies seen by the grain, the effects of which are maximal when $\nu_{\rm s} \sim \Omega$. Thus in principle, the Fourier transform of gas in the frame of the grain $|\tilde{\gvel}|^2$ can depend upon grain properties.} Thus, we have a transfer function that describes how fluctuations in the gas velocity translate into drift velocity fluctuations for neutral grains.

Performing an integral over $\Omega$ should yield the standard deviation of the drift velocity distribution.\footnote{In performing this integral, it is important to note that we must use the fourier transform of the gas velocity with respect to frequency, $|\gvel_\Omega|^2$, which is distinct from $|\gvel_k|^2$. For example, for $k^2|\gvel_k|^2 \propto k^{-5/3}$, $|\gvel_\Omega|^2 \propto \Omega^{-2}$.} Technically, we should integrate from an outer scale to an inner (viscous) scale, but we will make the approximation that the viscous scale is infinitely small. The main difference introduced by introducing a viscous scale is that the drift velocity fluctuations will scale more steeply with the stopping time when the stopping time (or Larmor time) of grains is smaller than the turnover time of a critically damped eddy \citep{Yan+Lazarian+Draine_2004}. 

We define the spectral index of the turbulence as follows. While the 3D velocity power spectrum is $\mathcal{P}({\bm k}) = |\tilde{\gvel}|^2$, by marginalizing over other directions, one can obtain parallel (to the mean magnetic field) $\mathcal{P}(k_{||})$, transverse $\mathcal{P}(k_\bot)$, and isotropic $\mathcal{P}(k)$ power spectra. For each of these, if we expect a power law $k^\alpha$, $\alpha$ is the spectral index. The transverse and isotropic cases include geometric factors of $k$, $k^2$ respectively, so that for e.g. \citet{goldreich1995toward} turbulence, the average over the azimuthal direction of $k|\tilde{\gvel}|^2$ is proportional to $k^{-5/3}$, while for the parallel direction, the average of $|\tilde{\gvel}|^2$ over the positive and negative directions is proportional to $k^{-2}$. 

For neutral grains and a spectral index of $\alpha=-5/3$, the aformentioned integral yields,
\begin{align}
    \frac{\sigma(\drift)^2}{\sigma(\gvel)^2} &= \frac{t_{\rm s}}{t_{\rm dyn}}{\rm arccot}{\left(\frac{t_{\rm s}}{t_{\rm dyn}}\right)}.\label{eq:analytic_drag_only}
\end{align}
Here, $t_{\rm dyn}$ is the turnover time of the largest eddies. Eq.~\ref{eq:analytic_drag_only} converges to 1 when $t_{\rm s}/t_{\rm dyn} \rightarrow \infty$ and otherwise becomes a power law with $\sigma(\drift)\propto t_{\rm s}^{1/2}$ as $t_{\rm s}\rightarrow 0$. In that limit, this expression is similar to that found in \citet{Draine+etal_1985} ($\sigma(\drift) = \sigma(\gvel) (t_{\rm s}/t_{\rm dyn})^{1/2}$) although that expression clearly breaks down in the limit of large Stokes number $t_{\rm s}/t_{\rm dyn} \gtrsim 1$ (cf. \S~\ref{sec:fpneutral}). It is also in rough agreement with the expressions found in \citet{cuzzi2003blowing}. However, their expressions are somewhat lower in magnitude and have a slower shift from the power law behavior to asymptotic behavior. 

The integral can also be evaluated analytically for spectral indices of $\alpha=-3/2$ and $\alpha=-2$. These slopes may correspond to the perpendicular and parallel directions of the \citet{boldyrev2005spectrum} theory. $\alpha=-2$ also corresponds to Burgers turbulence theory \citep{burgers1948mathematical}, as well as the findings from the compressible supersonic turbulence simulations of \citet{federrath2013universality}. The solutions for these slopes are more complex, so we do not write them here. They only differ in their asymptotic forms for $\sigma(\drift)$ as $t_{\rm s}/t_{\rm dyn}\ll 0$.
In particular, for $\alpha=-3/2$, in the asymptotic limit we have,
\begin{equation}
    \frac{\sigma(\drift)^2}{\sigma(\gvel)^2} \approx 1.21 \left(\frac{t_{\rm s}}{t_{\rm dyn}}\right)^{2/3}, \label{eq:neutralboldyrev}
\end{equation}
while for $\alpha=-2$, we have,
\begin{equation}
    \frac{\sigma(\drift)^2}{\sigma(\gvel)^2} \approx -2\ln\left(\frac{t_{\rm s}}{t_{\rm dyn}}\right)\left(\frac{t_{\rm s}}{t_{\rm dyn}}\right)^2.\label{eq:neutralburgers}
\end{equation}
This last expression differs only slightly from a linear scaling of $\sigma(\drift)$ with $t_{\rm s}$ or grain size.
Effectively, for $\alpha=-3/2$ ($-2$) larger (smaller) amplitude eddies are present with a turnover time approximately equal to the grain stopping time, and thus grains assume a larger (smaller) typical drift velocity.

Eq.~\ref{eq:analytic_drag_only} can be used to compute the standard deviation of dust velocities $\sigma(\dvel)$ as well. As in \citet{cuzzi2003blowing}, we can write $\sigma(\dvel)^2$ for the Kolmogorov $\alpha=-5/3$ spectrum as,
\begin{align}
    \frac{\sigma(\dvel)^2}{\sigma(\gvel)^2} &= 1-\frac{\sigma(\drift)^2}{\sigma(\gvel)^2} \nonumber \\
    &=1-\frac{t_{\rm s}}{t_{\rm dyn}}{\rm arccot}{\left(\frac{t_{\rm s}}{t_{\rm dyn}}\right)}.\label{eq:analytic_vel_drag_only}
\end{align}

\subsubsection{Charged grains}
This simple model can also be extended to include the impact of charge on the drift velocity distributions. In order to do so, we proceed as in \citet{moseley2023dust} and treat velocity components orthogonal to the mean magnetic field (which is along $\hat{\bm z}$) as complex numbers, e.g.
\begin{equation}
    \dvel_\bot \rightarrow v_x + i v_y.
\end{equation}
In this way, we can let the Lorentz force be represented by $i \omega_{\rm L}$ rather than a cross product. This only works if we assume that the field direction remains constant, e.g. the Alfv{\'e}n Mach number is very low. Regardless, this allows us to construct a simple model that works reasonably well in explaining the dust behavior.

Proceeding as before, we arrive at,
\begin{align}
    -i \Omega \tilde{\bm w} &= -(\nu_{\rm s} + i \omega_{\rm L})\tilde{\bm w} + i \Omega \tilde{\bm u},\\
    \implies |\tilde{{\bm w}}|^2 &= \frac{\Omega^2 t_{\rm s}^2}{1+(\Omega-\omega_{\rm L})^2t_{\rm s}^2}|\tilde{\bm u}|^2. \label{eq:charged_ho}
\end{align}
Now we see that there is an impact of resonance on the grain velocities. Adopting this form, we can evaluate $\sigma(\drift)^2$ for the common spectral indices as before. It should be noted that as differently polarized modes $\pm \Omega$ are no longer equivalent when $\omega_{\rm L} \neq 0$, we must average over both sets of modes, assuming an equal mix of right- and left-polarized modes. %For the indices, the $-2$ spectral index should not apply here, as we expect $\omega_{\rm L}$ not to impact motion in the direction parallel to the mean field. 
For $\alpha=-5/3$, we obtain a power law that is approximately independent of charge. In particular, so long as $\omega_{\rm L} t_{\rm dyn} > 1$,

\begin{equation}
\frac{\sigma(\drift)^2}{\sigma(\gvel)^2} \approx \frac{\pi}{2} \frac{t_{\rm s}}{t_{\rm dyn}}. \label{eq:driftmodel}
\end{equation}
Thus, for charged grains, we predict that the grain velocity is only limited by how weak the drag is. The reason there is no explicit $\omega_{\rm L}$ dependence is that while the amplitude at resonance of the filtering term in Eq.~\ref{eq:charged_ho} increases with $\omega_{\rm L}$, as $\omega_{\rm L}$ increases, the grains probe smaller and smaller eddies, and so the gas velocity dispersion at the relevant scale is smaller. This happens at exactly the rate so as to perfectly cancel out any $\omega_{\rm L}$ dependence. When $\omega_{\rm L} t_{\rm dyn} < 1$, the solution quickly reverts to that for neutral grains as $\omega_{\rm L} t_{\rm dyn}$ falls below $1$. As an example, for $\omega_{\rm L} t_{\rm dyn} = 0.8$, the solution only differs from the neutral solution by less than a factor of 2 as $t_{\rm s}/t_{\rm dyn} \rightarrow \infty$, and does not differ at all in the limit as $t_{\rm s}/t_{\rm dyn} \rightarrow 0$.

For $\alpha=-3/2$, there are essentially two regimes that follow different power law scalings. When $\omega_{\rm L}t_{\rm s} < 1$, the grains are essentially neutral, and drag forces dominate over Lorentz, so we revert to the $t_{\rm s}^{1/3}$ scaling of neutral grains in a $\alpha=-3/2$ power spectrum. For $\omega_{\rm L}t_{\rm s} > 1$, we have that,

\begin{align}
    \frac{\sigma(\drift)^2}{\sigma(\gvel)^2} \approx \frac{\pi}{3}\frac{t_{\rm s}}{t_{\rm dyn}}\left(\omega_{\rm L}t_{\rm dyn}\right)^{1/3}.
\end{align}
When $\omega_{\rm L} t_{\rm s}<1$, we revert to the neutral scaling Eq.~\ref{eq:neutralboldyrev}. As well, the unabated increase with $t_{\rm s}$ is only true so long as $\omega_{\rm L} t_{\rm dyn} > 1$.
Based on these calculations, we predict that for spectral indices between $\alpha=-3/2$ and $\alpha=-5/3$, there is only a very weak dependence of the drift velocity dispersion on the charge, with the dispersion scaling only at or below a $1/6$ power. For steeper slopes, an inverse charge dependence develops, with the grains tending to \textit{lower} r.m.s. drift velocities as the charge on grains increases. For example, for the $\alpha=-2$ slope, as long as $\omega_{\rm L} t_{\rm dyn}>1$ and $\omega_{\rm L} t_{\rm s} > 1$,
\begin{equation}
    \frac{\sigma(\drift)^2}{\sigma(\gvel)^2} \approx \pi\frac{t_{\rm s}}{t_{\rm dyn}}\left(\omega_{\rm L}t_{\rm dyn}\right)^{-1}.
\end{equation}
As with the $\alpha=-3/2$ law, when $\omega_{\rm L} t_{\rm s}<1$ or $\omega_{\rm L} t_{\rm dyn}<1$, we revert to the neutral scaling Eq.~\ref{eq:neutralburgers}.

Which direction the charge-to-mass ratio dependence goes is determined by competition between two effects. One effect is gyroresonance: the larger $\omega_{\rm L}$ is, the higher the amplitude of the velocity when driven at resonance. The other effect is that the gas fluctuations become smaller with increasing $\omega_{\rm L}$. The effect that dominates depends upon the slope of the power spectrum, with $\alpha=-5/3$ being the transition point.

When considering absolute dust velocities $\dvel$, we cannot employ the same \citet{cuzzi2003blowing} procedure where $\sigma(\gvel)^2 -\sigma(\drift)^2 = \sigma(\dvel)^2$. This only applies for neutral grains. For charged grains, it may be the case that $\sigma(\drift) > \sigma(\gvel)$, and so this relationship would imply negative values for $\sigma(\dvel)^2$. 

%However, we present the results for all three spectral indices in Figs.~\ref{fig:perp_scaling} and \ref{fig:para_scaling} in comparison to our numerical results.

\subsection{A simplified Fokker-Planck model}\label{sec:fp}
We use the time-independent Fokker-Planck equation to try to understand our equilibrium drift velocity probability density functions:
\begin{align}
    -\frac{\partial}{\partial w_j}(\mathcal{A}_j p({\drift})) + \frac{1}{2}\frac{\partial^2}{\partial w_j \partial w_k}(\mathcal{B}_{jk} p({\bm w})) &= 0.
\end{align}
Here, $p$ is the probability density function and $\mathcal{A}$ and $\mathcal{B}$ are the advection and diffusion coefficients. The coefficients are given by,
\begin{align}
    \mathcal{A}_j &\equiv \lim\limits_{\Delta t \rightarrow 0}\frac{1}{\Delta t}\int(w_j' - w_j)P({\bm w}',\Delta t | {\bm w})d^3{\bm w}' = \left( \frac{{\rm d}\langle \drift'\rangle}{{\rm d}t}\right)_{t=0},\label{eq:Aj}\\
    \mathcal{B}_{jk} &= \lim\limits_{\Delta t \rightarrow 0}\frac{1}{\Delta t}\int (w_j' - w_j)(w_k' - w_k)P({\bm w}',\Delta t | {\bm w})d^3{\bm w}'= \left( \frac{{\rm d}\sigma(\drift')^2}{{\rm d}t}\right)_{t=0}.\label{eq:Bjk}
\end{align}
Here, $P(\drift',\Delta t | \drift){\rm d}^3\drift'$ is the probability that a grain will go from a drift velocity $\drift \rightarrow \drift'$ in a time $\Delta t$ \citep{thorne2017modern}.

\subsubsection{Constant stopping time, neutral grain Fokker-Planck solution} \label{sec:fpneutral}

The term that drives the evolution of the velocity dispersion is the time derivative of the fluid velocity.
Integrate Eq.~\ref{eq:langevin} from $0$ to $\Delta t$. Then,
\begin{align}
    \Delta \driftvel = -\nu_{\rm s} \driftvel \Delta t - \Delta \gvel
\end{align}
Take the ensemble average and take the time derivative to find $\mathcal{A}$.
\begin{align}
    \mathcal{A} = \frac{{\rm d}\langle\Delta \driftvel\rangle}{{\rm d}\Delta t}\bigg|_{\Delta t = 0} = -\langle \nu_{\rm s}\rangle \driftvel. \label{eq:neutral_advection}
\end{align}
For the diffusion operator $\mathcal{B}$, we have in 1D,\footnote{Implicit in this expression is the assumption that the autocorrelation function $\langle u(t)u(t+\Delta t)\rangle = \sigma(\gvel)^2{\rm e}^{-\Delta t/t_{\rm dyn}}$. It is worth noting that this is a Lagrangian autocorrelation function taken in the frame of the particle, and so in general it may depend upon $t_{\rm s}$, though the dependence is weak for Stokes number $t_{\rm s}/t_{\rm dyn} \ll 1$. Different assumptions about this function are possible and will impact the final result \citep{cuzzi2003blowing}.}
\begin{align}
    \mathcal{B} = 2\sigma(\gvel)^2/t_{\rm dyn}. \label{eq:neutral_diffusion}
\end{align}

Assuming that the turbulence is isotropic and that the three components of velocity are independent of one another, the Fokker-Planck equation is,
\begin{align}
    \frac{\partial}{\partial w_j}(\nu_{\rm s} w_j p({\bm w})) +\frac{\partial^2}{\partial w_j^2}(t_{\rm dyn}^{-1}\sigma(\gvel)^2 p({\bm w})) = 0.\label{eq:FPdrift}
\end{align}
This simply defines a three-dimensional Ornstein-Uhlenbeck process, the solution to which is known to be a gaussian.
For a constant drag law, the mean of the distribution is zero and the variance is,
\begin{align}
    \sigma({\bm w})^2 = \frac{t_{\rm s}}{t_{\rm dyn}}\sigma(\gvel)^2. \label{eq:n0}
\end{align}
% The citation below isn't displaying as proceedings should. You should fix that.
This is again, the same expression as found in \citet{Draine+etal_1985} as well as \citet{cuzzi2003blowing} for their $n=1$ case in the limit of Reynolds number ${\rm Re}\rightarrow \infty$ and $t_{\rm s}/t_{\rm dyn} \rightarrow 0$. It is equivalent up to constants of order unity to our expressions for the drift velocity dispersion of grains for a $-5/3$ spectral index (Eqs.~\ref{eq:charged_ho},~\ref{eq:analytic_drag_only}) in either the limit where $t_{\rm s}/t_{\rm dyn} < 1$ or $\omega_{\rm L}t_{\rm dyn} > 1$.

In the case of an asymmetric distribution, Eq.~\ref{eq:n0} will hold for each independent axis. Note that for a small dust-to-gas mass ratio $\mu \ll 1$, Eq.~\ref{eq:n0} is equivalent to saying that equal amounts of energy per unit mass are dissipated by dust as by gas, $\sigma({\bm w})^2/t_{\rm s} \sim \sigma(\gvel)^2/t_{\rm dyn}$. This can also be inferred from the idea espoused by \citet{Yan+Lazarian+Draine_2004} that grains assume a velocity typical for eddies with a turnover time equal to the grain stopping time.

%Thus, as the Stokes number increases, the turbulence becomes less coherent in the frame of the dust. This incoherence limits the magnitude of the drift velocity.

\subsubsection{Accounting for the Lorentz force}
We have a stochastic differential equation of the form,
\begin{align}
    \dot{\driftvel}= {\bm M}(t)\cdot \driftvel - \dot{\gvel}(t),
\end{align}
where ${\bm M} = {\bm N} +{\bm \Omega}$ is a matrix with a symmetric part ${\bm N}$ (the drag) and anti-symmetric part ${\bm \Omega}$ (the Lorentz force). 

This can be solved analytically for a constant ${\bm M}$. The anti-symmetric portion of ${\bm M}$ ultimately does not contribute to the solution in this case, and in the end the asymmetry of the solution is only a function of the asymmetry of the turbulence. This can be thought of as the case of infinitely strong magnetic field. The solution will thus be an oblate gaussian. 

To get a less trivial result, assume that ${\bm N}$ is constant, but ${\bm \Omega}$ is not. It is not hard to show (using the same methods that applied to the neutral advection operator, Eq.~\ref{eq:neutral_advection}), that the coefficient $\mathcal{A}$ is given by,
\begin{align}
    \mathcal{A} = -\nu_{\rm s} \driftvel +\langle{\bm \Omega}\rangle\cdot \driftvel = -(\nu_{\rm s}+\langle \omega_{\rm L}\hat{\bm b}\rangle\,\times\,) \, \drift.
\end{align}
The second term here only serves to rotate the distribution around the mean field.

The diffusion coefficient $\mathcal{B}$ is more complex. We leave a more thorough explanation of its form in Appendix~\ref{sec:diffusion}. It is given by,
\begin{align}
    \mathcal{B} &= \left\langle {\bm \Omega}\cdot\driftvel \cdot \gvel^T - \gvel\cdot \driftvel^T\cdot {\bm \Omega}\right\rangle + 2\sigma(\gvel)^2t_{\rm dyn}^{-1}{\bm \delta}.\label{eq:diffusion_operator}
\end{align}
$T$ here denotes the transpose. It can be shown that the only relevant terms that emerge from the bracketed expression are proportional to the ensemble-averaged electric field along the guide magnetic field.\footnote{Showing this requires only that we assume rotational symmetry about the mean magnetic field direction.} Unfortunately, this would imply that so long as the ensemble-averaged electric field is zero, the solution will not differ from a triaxial gaussian distribution. It may be that going to third order (or higher) terms in the Fokker-Planck Equation is necessary, or it may be that the effect of intermittency on the grains is not properly accounted for in this theory.

Regardless, assume a form for the diffusion operator. Its terms are larger when $\driftvelmag$ is larger, owing to the terms linearly dependent on $\driftvelmag$ in Eq.~\ref{eq:diffusion_operator}. As well, note that the typical magnitude of ${\bm \Omega}$ is $\omega_{\rm L}$ and the typical value for $\gvel$ is $\sigma(\gvel)$. Suppose that the final distribution that emerges from this random process is isotropic. This means that, in spherical coordinates, any terms other than the purely radial terms are inconsequential, as velocity-space fluxes orthogonal to the radial direction are zero by virtue of isotropy. 

By inspection of Eq.~\ref{eq:diffusion_operator}, we provide a simple form that the radial portion of the diffusion operator may take,
\begin{align}
    \mathcal{B}_{\driftvelmag\driftvelmag} = 2\lambda \omega_L \sigma(\gvel) \driftvelmag,
\end{align}
where $\omega_L$ is the average Larmor frequency and $\lambda$ is between zero and one. $\lambda$ is introduced simply to account for the fact that there is some degree of correlation between ${\bf B}$ and $\gvel$ which is non-zero but also less than unity.

The Fokker-Planck equation for the drift velocity of grains with charge and a constant drag coefficient is then,
\begin{align}
    0=\frac{\partial}{\partial \driftvelmag}\left(\nu_{\rm s} \driftvelmag^3 p(\driftvelmag)\right) + \frac{\partial}{\partial\driftvelmag}\left(\driftvelmag^2\frac{\partial}{\partial\driftvelmag}\left(\left(\frac{\sigma(\gvel)^2}{t_{\rm dyn}} +\lambda\sigma(\gvel) \omega_L \driftvelmag\right)p(\driftvelmag)\right)\right).
\end{align}
the $p$ that appears here will need to be multiplied by $\driftvelmag^2$ in the end. 
This produces a solution where
\begin{align}
    &p(w)= \nonumber \\
    & \left(\frac{\lambda\omega_L t_{\rm dyn}}{\sigma(\gvel)}\right)^3\frac{n(n+1)}{n {\rm e}^n E(n)-1}\left(1+\frac{\lambda\omega_L t_{\rm dyn}}{\sigma(\gvel)} \driftvelmag\right)^{(n-1)}\driftvelmag^2 \exp\left(-\frac{\nu_{\rm s}\driftvelmag}{\lambda\omega_L\sigma(\gvel)}\right),\label{eq:fullsoln}\\
    %U_{-n}^{n}\left(\frac{\nu_{\rm s}\driftvelmag}{\omega_L\sigma(\gvel)}\right),\\
    &n\equiv \frac{\nu_{\rm s}}{\lambda^2\omega_L^2t_{\rm dyn}},\\
    &E(n) \equiv \int_{1}^\infty x^{n+1}{\rm e}^{-n x}{\rm d}x.
    %U_a^b(z)&\equiv \Gamma(a)^{-1}\int_0^\infty {\rm e}^{-z x} x^{a-1}(1+x)^{b-a-1}{\rm d}x 
\end{align}
%$U_a^b(z)$ is the confluent hypergeometric function and $\Gamma(a)$ is the $\Gamma$ function. % Question: Can I writ this purely in terms of n?
 This is very close to an ordinary exponential, especially when $\driftvelmag/\sigma(\gvel) \gg \omega_L t_s$. When $n\ll 1$ (strongly charged grains), we may write the above as
 \begin{align}
     &p(w)\approx \left(\frac{\nu_{\rm s}^2 t_{\rm dyn}}{\lambda\omega_L\sigma(\gvel)^3}\right)\frac{\driftvelmag^2}{1+\lambda\omega_L t_{\rm dyn} \driftvelmag/\sigma(\gvel)} \exp\left(-\frac{\nu_{\rm s}\driftvelmag}{\lambda\omega_L\sigma(\gvel)}\right).\label{eq:nll1}
 \end{align}
 In this case, it can be shown that the drift velocity dispersion scales as $t_{\rm s}^{1}$, whereas when $n\gg 1$ (weakly charged grains), we obtain $t_{\rm s}^{1/2}$. As well, the drift velocity dispersion scales as $\omega_{\rm L}^{1}$ when $n\lesssim1$ and $\omega_L^0$ when $n\gtrsim 1$.
 In the limiting case where $t_{\rm dyn} \to \infty$ (effectively ignoring the term from gas dispersion alone), we recover an exponential times $\driftvelmag^2$. Though it is not immediately obvious, Eq.~\ref{eq:fullsoln} also reduces to the constant-drag-coefficient gaussian profile in Eq.~\ref{eq:n0} in the limit where $\omega_L\to 0$. Naturally this also happens when we let the degree of correlation between the magnetic field and the fluid velocity $\lambda \to 0$.

 While this model may not be quantitatively correct, it offers an explanation for why introducing charge might cause grain drift velocity PDFs to go from gaussian to exponential in shape. In this picture, the reason is that when grains are charged, as they are accelerated to higher velocities, they diffuse more rapidly in velocity space, giving them a higher probability of diffusing to even higher velocities. Another way to see this is that if grains are accelerated upon entering a region with a particular field configuration, faster grains can sample these regions more often, and so can be accelerated even further.

% The diffusion operator can be simplified to show this framework says that there is no impact of the magnetic field except if there is a time-averaged parallel electric field. This suggests that the Fokker-Planck equation should be extended to higher moments for this analysis, rather than being cut off at the second moment.

\subsection{Summary of analytic models}
The above models, while non-rigorous, qualitatively explain the dust grain drift velocity probability density functions that we see in our simulations. Neutral grains follow a roughly gaussian profile whose width depends upon the square root of the grain size (i.e. strength of the drag) and the turbulent velocity dispersion $\sigma(\gvel)^2$ when $t_{\rm s}$ is much less than the turbulent turnover time $t_{\rm dyn}$, eventually turning over to $\sigma(\drift) \approx \sigma(\gvel)$ when $t_{\rm s} \gtrsim t_{\rm dyn}$. Charged grains have non-gaussian profiles because faster grains experience higher levels of diffusion, rather than all grains experiencing the same diffusion. Additionally, charged grains may, in principle, have both drift and absolute velocity distributions much wider than the gas velocity distributions when $t_{\rm s}/t_{\rm dyn} > 1$ and $\omega_{\rm L} t_{\rm dyn} > 1$. The width of these distributions should only be weakly dependent upon the charge according to the theories in Sec.~\ref{sec:qlt}, but linearly dependent in the theory presented in Sec.~\ref{sec:fp}. As we will see in the following section, both of these pictures help to explain our results. Drift velocity distribution widths are indeed only weakly charge dependent, validating the model in Sec.~\ref{sec:qlt}, while the shapes of the distribution are often strongly charge dependent, in line with our expectations from Sec.~\ref{sec:fp}.
%This happens because there is a cross-correlation between the fluid velocity and the magnetic field (or Lorentz force). 

%Grains that have a higher magnetization $\omega_L t_s$ have a broader velocity distribution, and so more readily reach high velocities. This means that magnetized grains are able to be accelerated to much higher velocities than unmagnetized grains. Thus, the properties of the Lorentz force in turbulence are critical to understanding the process of dust grain shattering in the ISM.

\section{Numerical Results} \label{sec:results}
\subsection{Root-mean-square grain velocities}
Fig.~\ref{fig:all_vels_vs_mach} shows a scatter plot of the root-mean-square grain drift and absolute velocities as a function of either perpendicular or parallel (to the mean field) gas turbulent Mach number for all of our grain populations after approximately eight turnover times. The simulations that these are drawn from are summarized in Tab.~\ref{tab:sims}. For both drift and absolute velocities, grains exhibit the baseline behavior that their velocities scale linearly with the gas velocity dispersion, with deviations from this trend occurring as a function of the grain properties, $\varepsilon$ and $\xi$. 

Grain-gas drift velocities exhibit substantially different behavior from absolute velocities. For absolute velocities, it is always the case that $\sigma(\dvel)\sim \sigma(\gvel)$, with deviations from this behavior never being more than perhaps $50\%$. This is not small by any means, but it is substantially less variation than is exhibited by the drift velocities, for which $\sigma(\drift)$ may be as much as an order of magnitude (or more) off from $\sigma(\gvel)$ for the parameters surveyed.\footnote{Note that we expect grains that are perfectly decoupled to have $\sigma(\drift) \sim \sigma(\gvel)$, and grains that are perfectly coupled to have $\sigma(\drift)\sim 0$.}
\subsubsection{Anisotropy}
Fig.~\ref{fig:asymmetry} shows the dependence of the anisotropy of the velocity distributions on the grain properties $\langle \omega_{\rm L} \rangle t_{\rm dyn}$ and $\langle \nu_{\rm s} \rangle t_{\rm dyn}$, as well as the turbulent Alfv{\'e}n Mach number $\mathcal{M}_{\rm A} = \sigma(\gvel)/v_{{\rm A},0}$, where $v_{{\rm A},0} \equiv |\langle {\bm B} \rangle|/\sqrt{4\pi \rho_0}$. The grains follow the intuitive pattern that when $\mathcal{M}_{\rm A} \gg 1$, the grain distributions become isotropic, because the turbulence itself is isotropic. On the other hand, when $\mathcal{M}_{\rm A} \ll 1$, the distributions may be highly anisotropic, with some simulations showing order of magnitude asymmetries in the grain populations. For the sub-Alfv{\'e}nic simulations, the anisotropy tends to increase with the charge-to-mass ratio parameter $\xi$ as well as the grain size $\varepsilon$.\footnote{$\langle \nu_{\rm s}\rangle t_{\rm dyn}\propto \varepsilon^{-1}$.} This is because as $\varepsilon$ or $\xi$ increase, the Lorentz force becomes more dominant, building upon anisotropy already present from the ambient turbulence. Interestingly, there are a number of simulations/populations where the anisotropy in velocity is such that the transverse velocities are much greater than the velocities along the field, while for those same simulations/populations, the grain relative velocities transverse to the field are \textit{smaller} than the velocities along the field. This is because in the sub-Alfv{\'e}nic regime, the Lorentz force can provide additional coupling transverse to the field, but not along it, while at the same time, if $\sigma(v_\bot)\sim\sigma(u_\bot)$, standard MHD turbulence theories would predict that $\sigma(v_\bot)/(\sqrt{2}\sigma(v_{||})) > 1$.

\subsubsection{Comparison to analytic models}\label{sec:comp_to_an}

To understand better the dependence of grain drift velocities on the grain properties, Figs.~\ref{fig:wperp_vs_ts},~\ref{fig:wpara_vs_ts} show the dependence of either $\sigma(w_\bot)/\sigma(u_\bot)$ or $\sigma(w_{||})/\sigma(u_{||})$ on $1/(\langle \nu_{\rm s}\rangle t_\bot)$ or $1/(\langle \nu_{\rm s}\rangle t_{||})$, respectively, with $t_\bot$ and $t_{||}$ being the dynamical times $\ell_0/(\sqrt{2}\sigma(u_\bot))$ and $\ell_0/\sigma(u_{||})$. $\langle \nu_{\rm s}\rangle$ is an ensemble average over the stopping rates $\nu_{\rm s} \equiv t_{\rm s}^{-1}$ of all grains in the given population. Points are also colored by their respective charge-to-mass ratio parameters $\xi \equiv \omega_{\rm L}t_{\rm A}$ as per the legend and caption. We plot each simulation individually to better isolate dependencies.

These plots also show the analytic models from Sec.~\ref{sec:qlt}. We can see that within each simulation (with the possible exception of the sub-Alfv{\'e}nic simulations), grains follow approximately $\sigma(\dvel) \propto t_{\rm s}^{1/2}$. Looking at the super-/trans-Alfv{\'e}nic simulations in particular, it seems apparent that the trend is slightly steeper and more linear for higher $\xi$, while as $\xi$ decreases, the trend looks more similar to the neutral grain model, leveling out more as $t_{\rm s} \rightarrow t_{\rm dyn}$ or as $\sigma(\dvel) \rightarrow \sigma(\gvel)$. In order to see if the trends continue where $t_{\rm s} > t_{\rm dyn}$, we have run one additional simulation, b1m1L (cf. Tab.~\ref{tab:sims}) which differs from b1m1 only in that grains are ten times larger. We can see that the trend does continue to some degree in this case, but at less than the $1/2$ slope predicted in Sec.~\ref{sec:qlt}. Additionally, differences between grains based upon charge-to-mass ratio become more apparent in this regime for these parameters, with lower $\xi$ grains leveling off more than higher $\xi$ grains. This suggests that the model proposed in Sec.~\ref{sec:qlt} is at least qualitatively correct. 

% Discuss parallel vs perpendicular
We expect distributions in the direction along the field to more closely resemble those of neutral grains. That is, we expect gaussian distributions with a width given by Eq.~\ref{eq:analytic_drag_only}. For super- and trans-Alfv{\'e}nic simulations, we do not expect this to necessarily be true, due to the lack of a strong guide field. Indeed, for those simulations, the results look quite similar to the transverse case, with the possible exception of b0.1m1, for which the Alfv{\'e}n Mach number is a modest 0.32. In b0.1m1, there is no obvious charge-to-mass dependence, but still a visible size dependence. 

% Then discuss sub-Alfvenic regime.
For sub-Alfv{\'e}nic simulations, we suspect that for the transverse motions, the highest field strength case (b0.1m0.1) may suffer from several limiting factors, giving rise to the unusual and non-monotonic behavior with the grain parameters. We expect that the simulation b10m0.1 is more representative of the actual transverse grain motions, with low charge-to-mass ratio grains approximately following Eq.~\ref{eq:analytic_drag_only} and higher charge-to-mass ratio grains better coupled, and so decreasing in their velocity dispersion with $\xi$. This is what we predicted in Sec.~\ref{sec:qlt} for spectral indices steeper than $-5/3$. As these are low Mach number simulations, they are quite diffusive, which can impact their power spectra. Indeed, upon inspection, these simulations exhibit a very steep spectrum for the perpendicular fluid velocities, as steep as $-5$. This can explain the behavior we see here, with perpendicular motions suppressed heavily as $\omega_{\rm L}$ increases. 

Along the field, we observe an extremely weak dependence on all grain parameters for our sub-Alfv{\'e}nic simulations. The dependence on grain size is approximately a $1/6$ power, while there is no obvious $\xi$ dependence (as expected). Examining the power spectra for the parallel velocities, we find that it is indeed very shallow, as shallow as $-1$. For this power spectrum, we would predict an extremely shallow, logarithmic dependence of the grain velocity dispersion on grain size. 

%We believe the reason for this unusual behavior is that as $\mathcal{M}_{||} \sim 0.01$ and $\mathcal{M}_{\bot} \sim 0.1$, while we run the simulations for eight dynamical times, this really corresponds to eight \textit{transverse} dynamical times. Thus, not even a single dynamical time has elapsed in the parallel direction in these simulations. Unfortunately, it is computationally infeasible to run these simulations for the amount of time necessary to determine if this is indeed the case. 

%At the same time, in this low-Mach number regime, the parallel and perpendicular dynamics are relatively independent of one another, so we can still gain useful information about the perpendicular direction.

Trends in absolute velocities are more difficult to assess. Within each simulation, there may be trends, but they do not necessarily persist in a systematic way across simulations. We believe this to be due to the stochastic nature of the turbulence itself. While statistics on $\drift$ are clear and have little noise, those on $\dvel$ are less so. We thus reserve a full exploration of $\dvel$ and its dependence on the various parameters for future work.

\subsection{Grain drift velocity probability density functions}
Beyond the typical velocities of grains, their probability density functions (PDFs) are also worth examining. In Sec.~\ref{sec:fp}, we used the time-independent Fokker-Planck equation to solve for the equilibrium drift velocity distributions of grains. For neutral grains, the solution is a gaussian in the sub-sonic regime, although when grains have $w \gtrsim c_{\rm s}$ so that the supersonic Baines correction becomes relevant (Eq.~\ref{eq:drag}), the distribution steepens. We also predicted that the distributions for charged grains would be roughly exponential, because grains that are moving faster diffuse at a higher rate in phase-space. 

Figs.~\ref{fig:b10_perp}–\ref{fig:b1d-1_para} show the survival functions (SFs, the probability that grains have a velocity greater than a given value $\driftvelmag$) of the grains in our simulations after eight turbulent turnover times. We plot these rather than the PDFs because they show a mostly identical shape, but with reduced noise, easier interpretation, and the added benefit that the vertical axis can be kept the same for each simulation.

In order to quantify the shapes of the PDFs, we have used the kurtosis. Specifically, a gaussian has a kurtosis of $3$, kurtoses larger than $3$ indicate large tails, and kurtoses less than $3$ indicate less power in the tails than a gaussian. A plot showing the kurtoses of the drift and absolute dust velocities as a function of the grain parameters can be seen in Fig.~\ref{fig:kurtosis}.

Focusing on Fig.~\ref{fig:b10_perp} as an example for the direction transverse to the field, we can see that in the super-Alfv{\'e}nic regime, distributions become more gaussian as $\xi$ decreases and more exponential as $\xi$ increases. Additionally, we see that for a given $\xi$, the distributions become more exponential as we decrease $\varepsilon$ as well. In the trans-Alfv{\'e}nic regime, we see only weak charge dependence, but strong grain size dependence for the distributions, with a roughly gaussian distribution for large $\varepsilon$ and exponential as $\varepsilon$ decreases. Finally, in the sub-Alfv{\'e}nic regime, the PDF width is of course a strong function of the grain charge (cf. Sec.~\ref{sec:comp_to_an}, Fig.~\ref{fig:wperp_vs_ts}), and the shape is more non-gaussian with larger, exponential tails for larger $\varepsilon$. 

In the direction along the field, we expect less charge dependence for sub-Alfv{\'e}nic simulations (Fig.~\ref{fig:b10_para}). In the super-Alfvenic regime, the PDFs look virtually identical to those in the transverse direction, as expected. For transsonic PDFs, we observe a similar patter, which perhaps somewhat less charge dependence than in the transverse direction. For the sub-Alfv{\'e}nic regime, we see that there is no charge dependence, and distributions are very close to gaussian. An additional sub-Alfv{\'e}nic case can be seen in Fig.~\ref{fig:b1d-1_para} in the middle row, corresponding to the simulation b0.1m1 (Tab.~\ref{tab:sims}). For these simulations, it seems that grain distributions are a weak function of all the relevant parameters, especially deep into the sub-Alfv{\'e}nic regime. In the mildly sub-Alfv{\'e}nic regime, it seems that in the parallel direction, distributions simply become more exponential as $\varepsilon$ decreases and more gaussian as $\varepsilon$ increases. 

In transsonic and supersonic simulations where the gas is either moderately or highly compressible, the drift velocity PDF of grains becomes more non-gaussian with higher outliers as the grain size shrinks. In \citet{moseley2023dust}, we showed that the grain drift velocity magnitude PDF is to some extent a function of the gas density. This is because of the density dependence of the stopping time (Eq.~\ref{eq:drag}). For a constant density, and for neutral grains, the theory in Sec.~\ref{sec:fp} that predicts gaussian distributions seems reasonable. However, for compressible simulations, the dust drift velocity PDF at a given density is essentially convolved with the gas density PDF. If this PDF is sufficiently wide, the shape of the drift velocity PDF will be altered. When a grain enters a region of dense gas, its drift velocity quickly decreases, and it also as a result spends a disproportionate amount of time in that region, getting ``stuck''. The density-dependent nature of the stopping rate thus distorts the PDF, causing a sharper core and wider (relatively speaking) tails.

Another explanation is that smaller particles couple to the Kolmogorov scale, producing wide, exponential tails \citep{pan2010relative, hubbard2012turbulence}. % Wang et al. 2000

In the other extreme, when grains are large enough that dense clumps have a spatial extent smaller than the grain stopping length, the grains effectively encounter a density of gas closer to the mean. The extreme density fluctuations in the gas are effectively averaged away from the point of view of these grains, and so they are able to maintain a more gaussian profile. 

The gas density dependence of the stopping time could also confer an additional dependence of $\sigma(\drift)/\sigma(\gvel)$ on the turbulent Mach number. This may be one reason that our highly supersonic simulations seem to show relatively higher $\sigma(\drift)/\sigma(\gvel)$ than our transsonic and (to a lesser extent) sub-Alfv{\'e}nic simulations.

For sub-sonic, incompressible simulations, we can see that the PDF shapes (cf. Fig.~\ref{fig:kurtosis}) are more strongly dependent upon grain charge than grain size. This supports the theory that it is the compressibility of the turbulence that is causing departure from a normal distribution. 

%Why across all regimes, $\varepsilon$ has an impact on the distribution shape in the transverse direction is unclear to us.
% Kurtosis is a good way to quantify how gaussian things are. Also, you need one figure that shows how well your trends fit. Just show different Omega_L, and sigma(w) vs ts against the different profiles. A 3x3 grid should work. You can choose to make it all in one plot or to split it up. 

\section{Discussion}\label{sec:discussion}
% We also must discuss the relative unimportance of thermal velocities of dust grains
As seen in Figs.~\ref{fig:wperp_vs_ts} and~\ref{fig:wpara_vs_ts}, the model for the r.m.s. drift velocities of charged dust grains presented in Eq.~\ref{eq:driftmodel} (dashed line in the figures) agrees reasonably well with our simulation results. The exception to this is those simulations that are significantly subsonic, and in that case, the discrepancy can be explained by the unusually steep power spectra. In this case, we would predict a strong charge dependence for the r.m.s. drift velocities such that grains with greater charge-to-mass ratios are better coupled to the field; this is exactly what we see. In the other simulations (trans- and supersonic), where the power spectra are closer to Kolmogorov, we observe an extremely weak charge dependence, and a power law in $t_{\rm s}/t_{\rm dyn}$ with a 1/2 slope, just as predicted in Eq.~\ref{eq:driftmodel}. The normalization appears hard to predict, although the model seems to consistently underpredict velocities in the highly supersonic (and super-Alfv{\'e}nic) limit, while getting them approximately correct in the transsonic limit. 

Building upon Eq.~\ref{eq:driftmodel}, we can give some estimates for the drift velocities of charged grains in the ISM. For simplicity, we ignore the supersonic correction when estimating the stopping time $t_{\rm s}$ of grains. This will lead to us under-predicting $t_{\rm s}$ by some amount; but as seen in Figs.~\ref{fig:wperp_vs_ts} and~\ref{fig:wpara_vs_ts}, we see that this is not necessarily a bad thing, as the theory previously underpredicted grain velocities in most cases. Assuming a mean molecular weight of 1.5, we have that in the cold neutral medium, the drift velocities of grains are approximately given by,

\begin{align}
    &\frac{\sigma(\drift)}{\sigma(\gvel)} \approx 1.25\,\left(\frac{t_{\rm s}}{t_{\rm dyn}}\right)^{1/2}\label{eq:dimensionless_number}\\
    &\frac{\sigma(\drift)}{\sigma(\gvel)} \approx 0.4 \,\times \nonumber\\
    &\left(\frac{\rho_{\rm d}^{\rm i}}{2\,{\rm g}\,{\rm cm}^{-3}}\right)^{1/2}\left(\frac{a_{\rm gr}}{0.1\,\mu{\rm m}}\right)^{1/2}\left(\frac{n_{\rm H}}{30\,{\rm cm}^{-3}}\right)^{-1/2}\left(\frac{c_{\rm s}}{{\rm km}\,{\rm s}^{-1}}\right)^{-1/2} \left(\frac{\ell}{1\,{\rm pc}}\right)^{-1/4}, \nonumber \\
    &\sigma(\drift) \approx 0.5 \,{\rm km}\,{\rm s}^{-1}\,\times \nonumber \\
    &\left(\frac{\rho_{\rm d}^{\rm i}}{2\,{\rm g}\,{\rm cm}^{-3}}\right)^{1/2}\left(\frac{a_{\rm gr}}{0.1\,\mu{\rm m}}\right)^{1/2}\left(\frac{n_{\rm H}}{30\,{\rm cm}^{-3}}\right)^{-1/2}\left(\frac{c_{\rm s}}{{\rm km}\,{\rm s}^{-1}}\right)^{-1/2} \left(\frac{\ell}{1\,{\rm pc}}\right)^{1/4}\label{eq:decoupling}
\end{align} % Also compare to Larson law. Or do this one as Kolmogorov, then the other as Larson. For Kolmogorov, you can just choose a 10 pc CNM cloud. The -1/3 vs. +1/10 dependence on scale depending upon the assumptions is worth including. What else?
$t_{\rm s}/t_{\rm dyn}$ is the Stokes number. Here, we have assumed the size-linewidth relation of \citet{Solomon+Rivolo+Barrett+Yahil_1987}. Eq.~\ref{eq:decoupling} quantifies the degree of dust-gas decoupling in velocity space at a given density, for a given grain size, grain composition, sound speed (effectively, temperature), and length scale. The fact that the dependence on the scale $\ell$ is so moderate suggests that $0.1\,\mu$m dust grains are only moderately coupled across a wide range of scales, and that the more important factors are the grain size and the gas density.\footnote{Assuming a Kolmogorov cascade rather than \citet{Solomon+Rivolo+Barrett+Yahil_1987}'s size-linewidth relation will instead yield a $-1/3$ slope, which is still relatively moderate.} 

There is some ambiguity here in whether we are discussing the velocity dispersion of grains in a cloud of size $L$ or we are examining the velocity dispersion of grains \textit{at a particular scale} $\ell$ within a larger cloud of density $n_{\rm H}$ and size $L>\ell$. If the former is meant, then we may use the \citet{Larson_1981} relations for cloud velocity dispersion and density as a function of scale to find that,
\begin{align}
    &\frac{\sigma(\drift)}{\sigma(\gvel)} \approx 0.06\left(\frac{\rho_{\rm d}^{\rm i}}{2\,{\rm g}\,{\rm cm}^{-3}}\right)^{1/2}\left(\frac{a_{\rm gr}}{0.1\,\mu{\rm m}}\right)^{1/2}\left(\frac{c_{\rm s}}{0.1\,{\rm km}\,{\rm s}^{-1}}\right)^{-1/2} \left(\frac{L}{1\,{\rm pc}}\right)^{1/10}, \nonumber \\
    &\sigma(\drift) \approx 0.06 \, {\rm km}\,{\rm s}^{-1}\times \nonumber\\ &\left(\frac{\rho_{\rm d}^{\rm i}}{2\,{\rm g}\,{\rm cm}^{-3}}\right)^{1/2}\left(\frac{a_{\rm gr}}{0.1\,\mu{\rm m}}\right)^{1/2}\left(\frac{c_{\rm s}}{0.1\,{\rm km}\,{\rm s}^{-1}}\right)^{-1/2} \left(\frac{L}{1\,{\rm pc}}\right)^{3/5}.\label{eq:cloudscale}
\end{align}

We see several interesting facts revealed here that were not previously obvious. First of all, as the dependence on cloud-scale $L$ is so shallow (just 1/10 for the ratio of $\sigma(\drift)$ to $\sigma(\gvel)$), we can expect the degree of gas-grain coupling to be roughly constant across clouds, and we expect grains to mostly remain within clouds. We also note how this differs from Eq.~\ref{eq:decoupling} which moreso applies \textit{within} clouds. Specifically, we predict that on sub-parsec scales within diffuse clouds, large grains are only marginally coupled to the gas, especially along field lines. On the other hand, when we consider the way that grains are coupled to clouds of a given size (and now by implication density), we find that grains are coupled to the overall cloud structure. Thus, grains may roughly follow the overall density structure of the ISM while still being marginally coupled in velocity-space.

This point may be a point of confusion, as Eq.~\ref{eq:cloudscale} predicts grains are well-coupled to \textit{whole clouds}, while Eq.~\ref{eq:decoupling} predicts that grains are only moderately coupled on to gas \textit{within clouds} on small scales. These relations also seem to suggest that grains are marginally coupled to the gas in more diffuse media and at intermediate and small scales, but strongly coupled to gas in dense clouds and on larger scales. This could suggest a tendency of grains to move from low densities to high densities, whereby they become ``stuck'' once they reach those higher density environments.

Also note that grains are better coupled to hotter clouds and more poorly coupled to colder clouds. This makes sense given the prevalence of shocks in the colder clouds, and the significant decoupling that those confer. It is also apparent that the size of the dust grain is quite important to the overall degree of coupling. % Is this really the explanation?
% Must think harder about this. Or, realistically, need to just summarize the important papers and move on... this paper should be finished within one day, or it can be.

We would also like to discuss our work in the context of \citet{commerccon2023dynamics}. In particular, the exponential shape of the drift velocity PDFs (and survival functions), which we hypothesize could be due to dust getting ``trapped'' in high density regions, could be partially of numerical origin, rather than physical origin. This would lead to a moderate suppression of the gas-grain r.m.s. velocity, in which case our results are in fact, more conservative, and on the side of greater dust-gas coupling (in velocity space), rather than less. While \citet{commerccon2023dynamics} discuss the nature of dust ``decoupling'' from gas, this is done so in the context of dust-to-gas mass ratio variations, a topic which we reserve for future work. % the fastest grains are almost certainly in the low density regions, (Moseley 2023), so note that....

Although they did not examine dust velocity distributions, it is also worth noting that the simulations of \citet{lee2017dynamics} are the most similar to our own across the literature. They model charged dust grains as a fully kinetic population of particles, although the fluid is treated with a Lagrangian method. However, like \citet{commerccon2023dynamics}, they focused on the dust-to-gas mass ratio fluctuations, rather than the velocity distribution. 

It is important to note that while the gas-grain drift velocity dispersion may be an appreciable fraction of the turbulent gas velocity dispersion, this will not \textit{necessarily} be reflected in the dust density. In particular, when $\omega_{\rm L}t_{\rm s}, \,\omega_{\rm L}t_{\rm dyn} > 1$, dust undergoes significant gyromotion about the center-of-mass velocity, which is most often approximately the gas velocity. This will suppress large-scale dust density variations across the mean field, even when the dust velocity is quite high. This becomes less true when the Larmor radius of grains increases, and when the drag force is larger. When there is significant gas motion (e.g. in shocks) on the scale of the dust Larmor radius, dust density fluctuations may be driven with a scale of order the dust Larmor radius \citep{moseley2023dust}. With a spectrum of sizes, this could be reflected as small-scale size sorting of grains, potentially impacting dust collisional processes in shocks and post-shock regions \citep{guillet2007shocks}.

\subsection{Implications for galaxy scale simulations}
In simulations with high spatial resolution (parsec or below, cf. Eq.~\ref{eq:decoupling}), dust grains may have appreciable motion relative to the gas, and so both diffuse and develop an appreciable pressure due to small-scale random motion.\footnote{This is not an ordinary, isotropic  diffusion, and indeed, dust may sometimes behave anti-diffusively, concentrating into sharp filaments.} For situations where one would expect that significant sub-grid turbulence would result in $\sigma(\drift) \gtrsim 0.1 \sigma(\gvel)$ at the grid scale, the assumption that dust is strictly coupled to gas is no longer warranted. While this assumption may produce accurate results in denser portions of the simulation, the transport and evolution of dust in the more diffuse media will not be accurate. Throughout these lower density regions, there may be significant diffusion of grains along magnetic field lines especially. In this way, our results may help to inform models and methods for dust in larger scale galactic simulations such as \citet{mckinnon2018simulating}, \citet{choban2022galactic}, and \citet{dubois2024galaxies}. Even in regions of these simulations where dust should in principle be well-coupled to the underlying gas, understanding the velocity distributions is important to accurately modeling the evolution of dust grains when considering dust-dust collisions, accretion of metals onto grains, and non-thermal sputtering.

\subsection{Caveats}
We have neglected a number of non-ideal effects, which in principle are important at the scale of the grain Larmor radius. For example, in CNM conditions, the ambipolar diffusion Reynolds Number is approximately 1 at a scale of 0.1 pc. This is approximately the same scale as the Larmor radius of a 1 $\mu$m dust grain. Thus, in principle non-ideal effects may significantly impact our results. Illustrating this fact, \citet{guillet2007shocks} examined the dynamics of dust in J- and C-type shocks in order to understand how grains may be processed in shocks. They found that the dynamics of dust in shocks varies significantly based upon $\omega_{\rm L} t_{\rm s}$, the grain size, and the shock type, impacting the level of grain-processing that occurs in these post-shock regions. In particular, $\omega_{\rm L}t_{\rm s}$ impacts whether the grains are primarily coupled to the neutral or charged species, while the ratio of the Larmor radius to the shock thickness will determined whether or not the grains stay coupled to the fluid, or undergo significant shock acceleration. In this sense, our simulations represent regimes where the ambipolar diffusion scale is small relative to the grain Larmor radius. Including ambipolar diffusion (or other non-ideal effects) will result in smoother fluid properties, and thus grains should be better coupled to the fluid in the presence of these effects. 

Another fact is that, while still an interesting measure of dust-gas coupling, the gas-grain drift velocity does not directly translate into the degree of spatial coupling. This is because dust grains undergo gyromotions; therefore, they diffuse more along the magnetic field than across it, even at a given velocity dispersion. Given the drag that grains experience, they will still diffuse both across and along field lines, but for grains where $\omega_{\rm L} t_{\rm s} > 1$, they will do so slowly. We might thus expect dust density structures to form along the magnetic field. We hope to explore this topic in future work. 

%\subsection{Connections to other methods}
%On larger, cosmological scales, Dubois, Rodriguez, Vogelsberger, Choban.

%\subsection{}

%Implications for sputtering, grain shattering, etc.
% Scaling to physical units

% Resolution, driving procedure effects
% 
%Compare to Cuzzi, YLD and their later work.

% Also discuss in relationship to the work on neutral grain dispersion

%Also don't forget Ormel, Commercon, Windmark

% Then also the context for what these numbers mean in the context of various environments
% Can look at Hopkins papers for environment ideas... 
\section{Summary \& Conclusions}\label{sec:summary}
In this paper, we have systematically examined the way that the gas-grain relative (drift) velocity varies with turbulence and grain parameters. We have presented the results from 10 simulations, each with 16 different grain populations, varying the strength of the magnetic field, the strength of the turbulence, the grain charge, and the grain size. Each of these is quantified with a dimensionless number for generality. 

We presented simple theoretical models for the gas-grain velocity distribution and its standard deviation. One of these models was a 1D time-domain quasi-linear model, while the other was a Fokker-Planck model for the overall PDF. The 1D model explains well the r.m.s. drift velocity of grains, while the Fokker-Planck model is useful for understanding the shape of the PDF. We then compared these models to our simulation data. Drift velocity distributions in the simulations vary from almost exactly gaussian, to exponential, to distributions with extremely high kurtosis ($\approx 100$), depending upon both the grain size, the charge-to-mass ratio of grains, and the compressibility of the turbulence. We find that the kurtosis increases with grain charge and decreases with grain size. We also find that the distributions of velocity components parallel to the mean field depend more weakly on grain charge than across the field.

The dependence of the r.m.s. drift velocity on charge is weak, while the dependence on grain size is stronger. The grain size dependence, to factors within order unity, follows a simple power law with $\sigma(\drift)\approx 1.25\, {\rm St}^{1/2} \sigma(\gvel)$. This power law bends over (giving $\sigma(\drift)\sim \sigma(\gvel)$) at a Stokes number of $1$ for grains which are neutral or nearly neutral, and appears to continue for charged grains. Thus, in the high-Stokes regime, the grain population may have more kinetic energy per unit mass than the gas. This power law is equivalent to saying that equal energy per unit mass is dissipated by the dust as by the gas, $\sigma(\drift)^2/t_{\rm s} \sim \sigma(\gvel)^2/t_{\rm dyn}$.

We then discussed how our findings translate into the real interstellar medium. We found that typical 0.1 $\mu$m grains can be expected to drift at around 40\% of the gas velocity or more on parsec and smaller scales in the CNM. This translates to 0.5\kms or so, although the velocity dispersion of the gas may vary and will affect this significantly. On the scales of whole clouds (or denser media), we found that grains are much more strongly coupled, having a drift of just 6\% or so of the gas velocity dispersion. This implies that grains are well coupled to overall clouds, even while on smaller scales or in less dense media they may only be marginally coupled. 

Finally, we outlined one potential future application of this work to galactic scale grain evolution models, both in terms of describing an effective ``dust pressure'' and in terms of grain size evolution through shattering and coagulation. We then addressed a few caveats, most notably the fact that we have neglected non-ideal effects, which could have an important impact on grain dynamics.

%The question of ``how coupled'' grains are to gas is nuanced, and depends upon whether one means spatial coupling or velocity coupling.
%Distributions of neutral or near-neutral grains are gaussian, while strongly charged grains may follow either a gaussian or exponential distribution, depending upon the strength of the drag. 
% scaling in different regimes of the RMS
% relative vs absolute
%Follow up work
% Context

%\section{Summary}

\section*{Data availability}
The data presented in this paper was generated on the Princeton computing cluster `Stellar' and will be made freely available upon request to the corresponding author.
\section*{Acknowledgements}
We would like to thank B. T. Draine for comments which greatly improved the paper. We would also like to thank Ulrich Steinwandel for helpful discussions. 

\appendix 
\section{Further details of the Fokker-Planck analysis}
\subsection{The diffusion operator} \label{sec:diffusion}
For the diffusion operator $\mathcal{B}$ of neutral grains, we have in 1D,
\begin{align}
    \frac{{\rm d}\langle \Delta \driftvelmag^2\rangle}{{\rm d}\Delta t}\bigg|_{\Delta t=0} &= \frac{{\rm d}}{{\rm d}\Delta t}\bigg|_{\Delta t=0}\langle \Delta u^2\rangle,\\
    &= \frac{{\rm d}}{{\rm d}\Delta t}\bigg|_{\Delta t=0}\langle u(t+\Delta t)^2 -2u(t)u(t+\Delta t)+u(t)^2\rangle,\\
    &= \frac{{\rm d}}{{\rm d}\Delta t}\bigg|_{\Delta t=0}\left(2\sigma(\gvel)^2(1-{\rm e}^{-\Delta t/t_{\rm dyn}})\right),\\
    \implies \mathcal{B} &= 2\sigma(\gvel)^2/t_{\rm dyn}.
\end{align}
In the last step, we have assumed that the autocorrelation function $\langle u(t)u(t+\Delta t)\rangle = \sigma(\gvel)^2{\rm e}^{-\Delta t/t_{\rm dyn}}$. It is worth noting that this is a Lagrangian autocorrelation function taken in the frame of the particle, and so in general it may depend upon $t_{\rm s}$, though the dependence is weak for Stokes number $t_{\rm s}/t_{\rm dyn} \ll 1$. Different assumptions about this function are possible and will impact the final result \citep{cuzzi2003blowing}.

However, for neutral grains, when $t_{\rm s}/ t_{\rm dyn} \gtrsim 1$, the validity of this expression becomes questionable, although to what extent it becomes invalid is a question of the nature of the random process generated by the turbulent flow. Comparing the diffusion operator's form and the solution in Eq.~\ref{eq:analytic_drag_only}, we may suppose that the Stokes-number dependant nature of the autocorrelation function (and thus the diffusion operator) might take the form,
\begin{align}
    {\rm e}^{-\Delta t/t_{\rm dyn}} \rightarrow {\rm e}^{-\Delta t/t_{\rm dyn}} {\rm arccot}\left(\frac{t_{\rm s}}{t_{\rm dyn}}\right),
\end{align}
or a similar expression.\footnote{This expression does not actually have the correct normalization for an autocorrelation function.}

For charged grains, we may write the diffusion coefficient as,
\begin{align}
    \mathcal{B} &= \frac{{\rm d}}{{\rm d}\Delta t}\bigg|_{\Delta t=0}\langle \Delta \driftvel \cdot \Delta \driftvel^T\rangle,\\
    &= \frac{{\rm d}}{{\rm d}\Delta t}\bigg|_{\Delta t=0} \nonumber\\
    &\bigg\langle-\int_t^{t+\Delta t}{\bm \Omega}(\tau){\rm d}\tau\cdot\driftvel\cdot\Delta \gvel^T + \Delta \gvel\cdot \driftvel^T\cdot\int_t^{t+\Delta t}{\bm \Omega}(\tau){\rm d}\tau +\nonumber\\
    &\Delta \gvel\cdot\Delta\gvel^T - \int_t^{t+\Delta t}{\bm \Omega}(\tau){\rm d}\tau\cdot \driftvel \cdot\driftvel^T\cdot\int_t^{t+\Delta t}{\bm \Omega}(\tau){\rm d}\tau\bigg\rangle.
\end{align}
Note that we have neglected any terms that appear due to the drag. That is because either the ensemble average of those terms will be zero, or it will not depend upon time, and so the time derivative will eliminate it. Addressing the last two terms, we know what the product of the $\Delta \gvel$ terms produces. It will simply contribute $2\sigma(\gvel)^2t_{\rm dyn}^{-1}{\bm \delta}$. The last term is from the action of the random matrix ${\bm \Omega}$ alone. Its derivative with respect to $\Delta t$ is zero when $\Delta t = 0$, so we can neglect it as well. 

Next, the terms involving $\gvel(t+\Delta t)$ and the integral of ${\bm \Omega}$ will be zero, as their ensemble average will be ${\Delta t}$ independent. That leaves only those terms $\gvel(t)$. Passing the derivative into the ensemble average gives the result in Eq.~\ref{eq:diffusion_operator}.

\subsection{Including the Kwok correction in 1D}
Allow for a the more complex Epstein-Baines drag law as in Eq.~\ref{eq:drag}. For simplicity, write it as,
\begin{align}
    \nu_{\rm s} &= \nu_{{\rm s},0}\bigg(1+\frac{w^2}{a^2}\bigg)^{1/2}. 
\end{align}
Here, $a = c_{\rm s} \sqrt{128/(9\pi \gamma)}$. 

Now we again solve Eq.~\ref{eq:FPdrift}, but allowing for a velocity dependent stopping time. The solution is,
\begin{align}
    p({\bm w})\propto \exp\bigg[-\frac{\nu_{{\rm s},0}t_{\rm dyn}}{3\sigma(\gvel)^2/a^2}(1+w^2/a^2)^{3/2}\bigg].
\end{align}
The normalization does not seem to admit an analytic solution. 

For small $w \ll a$, this distribution is simply a gaussian.
\begin{align}
    p({\bm w})\propto \exp\bigg[-\frac{\nu_{{\rm s},0}t_{\rm dyn}}{2\sigma(\gvel)^2}w^2\bigg].
\end{align}
When $w \gg a$, the distribution steepens so that
\begin{align}
    p({\bm w}) \sim  \exp\bigg[-\frac{\nu_{{\rm s},0}t_{\rm dyn}}{3\sigma(\gvel)^2a}w^3\bigg].
\end{align}
The kurtosis of this distribution depends upon the ratio of $\sigma(\gvel)^2/(\nu_{{\rm s},0} t_{\rm dyn})$ to $a^2$. For large $\sigma(\gvel)^2/(\nu_{{\rm s},0} t_{\rm dyn})$, the kurtosis approaches zero, as power is drained more and more dramatically from the tails of the distribution in a relative sense. For very small $\sigma(\gvel)^2/(\nu_{{\rm s},0} t_{\rm dyn})$, the distribution approaches a kurtosis of 3, as it simply becomes a gaussian in this limit.

% TODAY:
% Add c_{\rm s} into the code (later...) Make final OT slice figure.
% Make planar shock figure. Remake RDI growth fig.
% Make Decaying turb figs:
% 1. Cubes at various times
% 2. Trajectories & accelerations
% 3. Histograms: dust vs gas density, dust-to-gas mass ratio (integrated vs volume)
% 4. Acceleration histograms
% 5. CDF of grain velocity vs gas velocity
% 6. Do the perturbation theory for a sound wave
% 7. High-k regime for hydro-RDI:
% k c_{\rm s} t_s > 1/\mu
% For \mu = 0.1, this means
% ts > 1.6
% Could just let ts=3.0 to be safe, but
% need to figure out the corresponding 
% grain size.

% Then you have all the figures! You're done.
%Follow-up papers:
% Driven turb (+AMR?) with grain size spectrum (analyze how far down the sub-structure goes?) and compare to single-size
% See impact of grains on the PS for single-size?
% Dust-to-gas mass ratio: monte-carlo method for  position updates
% Implement collisions somehow: 

%%%%%%%%%%%%%%%%%%%%%%%%%%%%%%%%%%%%%%%%%%%%%%%%%%

%%%%%%%%%%%%%%%%%%%% REFERENCES %%%%%%%%%%%%%%%%%%

% The best way to enter references is to use BibTeX:

%\bibliographystyle{mnras}
%\bibliography{example} % if your bibtex file is called example.bib

% Alternatively you could enter them by hand, like this:
% This method is tedious and prone to error if you have lots of references
\bibliographystyle{mnras}
\bibliography{ms_reduced}

%%%%%%%%%%%%%%%%%%%%%%%%%%%%%%%%%%%%%%%%%%%%%%%%%%
% Hello hello hello hello
%%%%%%%%%%%%%%%%% APPENDICES %%%%%%%%%%%%%%%%%%%%%

\appendix
\end{document}